\documentclass[11pt]{article}
\usepackage[hypertexnames=false]{hyperref}
\usepackage{color}
\usepackage[dvipsnames]{xcolor}
\usepackage{latexsym}
\usepackage{dsfont}
\usepackage{appendix}
\usepackage{amssymb}
\usepackage{subcaption}
\usepackage[margin=.9in]{geometry}
\usepackage{float}
\usepackage{algorithm,algorithmic}
\usepackage[numbers]{natbib}
\usepackage{amsmath, amsfonts,amssymb,euscript,array,enumerate,amsfonts,mathrsfs}
\usepackage[dvipsnames]{xcolor}
\usepackage{amsthm}
\newtheorem{Theorem}{Theorem}[section]

\usepackage[T1]{fontenc}
\usepackage[latin1]{inputenc}
\usepackage{hhline}
\usepackage{graphicx}
\usepackage{verbatim}
\usepackage{eurosym}
\usepackage{dsfont}
\allowdisplaybreaks

\long\def\ben#1{{\color{blue} #1}}

\DeclareMathOperator{\E}{\mathbb{E}}

\DeclareMathOperator{\R}{\mathbb{R}}
\DeclareMathOperator{\Sc}{\mathcal{S}}
\DeclareMathOperator{\Ac}{\mathcal{A}}
\DeclareMathOperator{\Pc}{\mathcal{P}}

\makeatletter
\@addtoreset{equation}{section}

\newcommand*{\rom}[1]{\expandafter\@slowromancap\romannumeral #1@}
\makeatother
\newcommand{\vertiii}[1]{{\left\vert\kern-0.25ex\left\vert\kern-0.25ex\left\vert #1 
    \right\vert\kern-0.25ex\right\vert\kern-0.25ex\right\vert}}

\begin{document}
\title{Recent Advances in Reinforcement Learning in Finance}

\author{Ben Hambly
\thanks{Mathematical Institute, University of Oxford. \textbf{Email:} \{hambly,  yang\}@maths.ox.ac.uk }
\and
Renyuan Xu \thanks{Epstein Department of Industrial and Systems Engineering, University of Southern California. \textbf{Email:} renyuanx@usc.edu}
\and
Huining Yang\thanks{ 
Supported by the EPSRC Centre for Doctoral Training in Industrially Focused Mathematical Modelling (EP/L015803/1) in collaboration with BP plc.}
  \footnotemark[1]
}
\maketitle

\begin{abstract}

The rapid changes in the finance industry due to the increasing amount of  data have revolutionized the techniques on data processing and data analysis and brought new theoretical and computational challenges.  In contrast to classical stochastic control theory and other analytical approaches for solving financial decision-making problems that heavily reply on model assumptions, new developments from reinforcement learning (RL) are able to make full use of the large amount of financial data with fewer model assumptions and to improve decisions in complex financial environments.
This survey paper aims to review the recent developments and use of RL approaches in finance. We give an introduction to Markov decision processes, which is the setting for many of the commonly used RL approaches. Various algorithms are then introduced with a focus  on value-based and policy-based methods that do not require any model assumptions. Connections are made with neural networks to extend the framework to encompass deep RL algorithms. We then discuss in detail the application of these RL algorithms in a variety of decision-making problems in finance, including optimal execution, portfolio optimization, option pricing and hedging, market making, smart order routing, and robo-advising. Our survey concludes by pointing out a few possible future directions for research.


\end{abstract}

\tableofcontents
\section{Introduction}

The mathematical approach to many financial decision-making problems has traditionally been through modelling with stochastic processes and using techniques from stochastic control. The choice of models is often dictated by the need to balance tractability with applicability. Simple models lead to tractable and implementable strategies in closed-form or that can be found through traditional numerical methods. However, these models sometimes oversimplify the mechanisms and the behaviour of financial markets which may result in strategies that are sub-optimal in practice and that can potentially result in financial losses. On the other hand, models that try to capture realistic features of financial markets are much more complex and are often mathematically and computationally intractable using the classical tools of stochastic optimal control.


In recent years the availability of large amounts of financial data on transactions,
quotes and order flows in electronic order-driven markets has
revolutionized data processing and statistical modeling techniques in finance and brought new theoretical and computational challenges \cite{dixon2020machine}. In contrast to the classical stochastic control approach, new ideas coming from reinforcement learning (RL) are being developed to make use of all this information. Reinforcement learning describes methods by which agents acting within some system might learn to make optimal decisions through repeated experience gained by interacting with the system.
In the finance industry there have been a number of recent successes in 
applying  RL algorithms in areas such as order execution, market making and portfolio optimization that have attracted a lot of attention. This has led to rapid progress in adapting RL techniques to improve trading decisions in various financial markets when participants have limited information on the market and other competitors.


Although there are already a number of more specialized review papers concerning aspects of reinforcement learning in finance, we aim to review a broad spectrum of activity in this area. This survey is intended to provide a systematic introduction to the theory of RL, provide a unified framework for performance evaluation and a comprehensive summary of the cutting-edge results in RL theory. {It is} followed by an introductory discussion of each of the following financial problems -- optimal execution, portfolio optimization, option pricing and hedging, market making, smart order routing, and robo-advising. Moreover, we will also discuss the advantages of RL methods over classical approaches such as stochastic control, especially for the problems that have been studied extensively in the mathematical finance literature. 
For other surveys in the literature on RL applications in finance, see \cite{charpentier2020,fischer2018,kolm2020modern,meng2019,mosavi2020}. The main focus of these surveys are traditional financial applications such as portfolio optimization and optimal hedging  \cite{charpentier2020,kolm2020modern}, or trading on the stock and foreign exchange markets \cite{meng2019}, or specific RL approaches such as actor-critic-based methods \cite{fischer2018} and deep RL methods \cite{mosavi2020}.  

Our survey will begin by discussing Markov decision processes (MDP), the framework for many reinforcement learning ideas in finance. 
We will then consider different approaches to learning within this framework with the main focus being on value-based and policy-based methods. In order to implement these approaches we will introduce deep reinforcement methods which incorporate deep learning ideas in this context. For  financial applications we will consider a range of topics and for each we will introduce the basic underlying models before considering the RL approach to tackling them. We will discuss a range of papers in each application area and give an indication of their contributions. We conclude with some thoughts about the direction of development of reinforcement learning in finance.

\section{The Basics of Reinforcement Learning}\label{sec:rl_basics}
Reinforcement learning is 
an approach to understanding and automating goal-directed learning and decision-making. It leads naturally to algorithms for such problems and is distinguished from other computational approaches by its emphasis on learning by the individual from direct interaction with {their} environment, without relying on exemplary supervision or complete models of the environment.
RL uses a formal framework defining the interaction between a learning agent and {their} environment in terms of states, actions, and rewards. This framework is intended to be a simple way of representing essential features of the artificial intelligence problem. These features include a sense of cause and effect, a sense of uncertainty and non-determinism, and the existence of explicit goals.

The history of RL has two main threads that were pursued independently before intertwining in modern RL. One thread concerns learning by trial and error and started in the psychology of animal learning. 
The other thread concerns the problem of optimal control for an evolving system and its solution using value functions and dynamic programming.  Although this thread did not involve learning directly, the Bellman equation developed from this line of literature is viewed as the foundation of many important modern RL algorithms such as $Q$-learning and Actor-Critic.

Over the last few years, RL,  grounded on combining classical theoretical results with deep learning and the functional approximation paradigm, has proved to be a fruitful approach to many artificial intelligence tasks from diverse domains. Breakthrough achievements include reaching human-level performance in such complex tasks as Go \cite{silver2017mastering} and multi-player StarCraft II \cite{vinyals2019grandmaster}, as well as the development of ChatGPT \cite{radford2019language}. The generality of the reinforcement learning framework allows its application in both discrete and
continuous spaces to solve tasks in both real and simulated environments \cite{lillicrap2015continuous}. 

Classical RL research during the last third of the previous century developed an extensive theoretical core on which modern algorithms are based. Several algorithms, such as Temporal-Difference Learning and $Q$-learning, were developed and are able to solve small-scale problems when either the states of the environment can be enumerated (and stored in memory) or the optimal policy lies in the space of linear or quadratic functions of the state variable. Although these restrictions are extremely limiting, these foundations of classical RL theory underlie the modern approaches which benefit from increased computational power. 
Combining this framework with deep learning \cite{goodfellow2016deep} was popularized by the Deep $Q$-learning algorithm, introduced in \cite{mnih2013playing}, which was able to play any of 57 Atari console games without tweaking the network architecture or algorithm hyperparameters. This novel approach was extensively researched and significantly improved over the following years \cite{van2016deep,gu2016continuous,fan2020theoretical}.

Our aim in this section is to introduce the foundations of reinforcement learning. We start with the setup for MDP in Section  \ref{sec:set_up} with both an infinite time horizon and a finite time horizon, as there are financial applications of both settings in the literature. In Section \ref{sec:MDP2learning} we then introduce the learning procedure when the reward and transition dynamics of the MDPs are unknown to the decision-maker. In particular, we introduce various criteria to measure the performance of learning algorithms in different settings. Then we focus on two major classes of model-free reinforcement learning algorithms, value-based methods in Section \ref{sec:value_based_methods} and policy-based methods in Section~\ref{subsec:policy_based_methods}, with an infinite-time horizon. Finally, Section~\ref{sec:discussion} contains a general discussion of (model-based and model-free) RL with a finite-time horizon, model-based RL with an infinite-time horizon, and regularization methods for RL. 

\subsection{Setup: Markov Decision Processes}\label{sec:set_up}

We first introduce MDPs with infinite time horizon. Many portfolio optimization problems, such as those arising from pension funds, investigate long-term investment strategies and hence it is natural to formulate them as a decision-making problem over an infinite time horizon. We then introduce MDPs with finite time horizon. Problems such as the optimal execution of the purchase or liquidation of a quantity of asset usually involve trading over a short time horizon and there may be a penalty at the terminal time if the targeted amount is not completely executed by then. Finally we discuss different policy classes that are commonly adopted by the decision-maker.

\paragraph{Infinite Time Horizon and Discounted Reward.} Let us start with a discrete time MDP with an infinite time horizon and discounted reward. We have a Markov process taking values in a state space $\Sc$, where we can influence the evolution by taking an action from a set $\Ac$. The aim is to optimize the expected discounted return from the system by choosing a policy, that is a sequence of actions through time. We formulate this mathematically by defining the  value function $V^\star$ for each $s\in \Sc$ to be 
 \begin{eqnarray} \label{equ: singlev}
V^\star(s) = \sup_{\Pi} V^{\Pi}(s):=
 \sup_{\Pi}\E^{\Pi}\biggl[\sum_{t=0}^{\infty}\gamma^t r(s_t, a_t) \biggl| s_0 = s\biggl],
 \end{eqnarray}
 subject to
 \begin{eqnarray} \label{equ: singledynamics}
 s_{t + 1} \sim P(s_t, a_t), \;\;\; a_t \sim \pi_t({s_t}).
 \end{eqnarray}
Here and throughout the paper, $\E^{\Pi}$ denotes the expectation under the policy ${\Pi}$, where the probability measure $\mathbb{P}$ describes both dynamics and the rewards in the MDP. We will write $\mathcal{P}(\mathcal{X})$ for the probability measures over a space $\mathcal{X}$. The state space $(\Sc, d_{\Sc})$ and the action space $(\Ac, d_{\Ac})$ are both complete separable metric spaces, which includes the case of  $\Sc$ and $\Ac$  being finite, as often seen in the RL literature; $\gamma$ $\in$ $(0, 1)$ is a  discount factor; $s_t \in \Sc$ and $a_t \in \Ac$ are the state and the action at time $t$; $P:  \Sc \times \Ac \to \Pc(\Sc)$ is the transition function of the underlying Markov process; the notation $ s_{t + 1} \sim P(s_t, a_t)$ denotes that $ s_{t + 1}$ is sampled from the distribution $ P(s_t, a_t)\in \mathcal{P}(\mathcal{S})$;  the reward  $r(s, a)$ is a random variable in $\R$ for each pair $(s, a)\in \Sc \times \Ac$; and the policy $\Pi =\{\pi_t\}_{t=0}^{\infty}$ is Markovian, in that it just depends on the current state, and can be either deterministic or randomized. For a deterministic policy, $\pi_t: \Sc \to \Ac$ maps the current state $s_t$ to a deterministic action. On the other hand, a randomized policy $\pi_t: \Sc\to \Pc(\Ac)$ maps the current state $s_t$ to a distribution over the action space.
For a randomized  policy $\pi_t$, we will use   $\pi_t(s)\in \mathcal{P}(\mathcal{A})$ and $\pi_t(s,a)$ to represent, respectively, the distribution over the action space given state $s$ and the probability of taking action $a$ at state $s$.
In this case, with infinite-time horizon, we assume the reward $r$ and the transition dynamics $P$ are time homogeneous, which is a standard assumption in the MDP literature \cite{puterman2014markov}. We also note that the setup is essentially the same if we consider the problem of minimizing costs rather than maximizing a reward.


The well-known Dynamic Programming Principle (DPP), that the optimal policy can be obtained by maximizing the reward from one step and then proceeding optimally from the new state, can be used to derive the following Bellman equation for the value function \eqref{equ: singlev};
\begin{eqnarray} \label{equ:classicalV}
V^\star(s) = \sup_{a \in \Ac} \Big\lbrace \E[r(s, a)] + \gamma \E_{s' \sim P(s, a)}[V^\star(s')] \Big\rbrace.
\end{eqnarray}
We write the value function as
$$V^\star(s) = \sup_{a \in \Ac} Q^\star(s, a),$$ 
where the $Q$-function, one of the basic quantities used for RL, is defined to be
\begin{eqnarray} \label{defsingleQ}
Q^\star(s, a) = \E[r(s, a)] + \gamma \E_{s' \sim P(s, a)}[V^\star(s')],
\end{eqnarray}
the expected reward from taking action $a$ at state $s$ and then following the optimal policy thereafter.
There is also a Bellman equation for the $Q$-function given by
\begin{eqnarray} \label{equ: singleQ}
Q^\star(s, a) = \E[r(s, a)] + \gamma \E_{s' \sim P(s, a)}\sup_{a' \in \Ac} Q^\star(s', a').
\end{eqnarray}
This allows us to retrieve the 
optimal (stationary) policy $\pi^*(s, a)$ (if it exists) from  $Q(s, a)$, in that  $\pi^*(s, a) \in \arg\max_{a \in \Ac} Q(s, a)$. 

An alternative setting over an infinite time horizon is the case of average reward, which is also referred to as an ergodic reward. This ergodic setting is not as relevant to financial applications and hence will not be covered in this review paper. We refer  the reader to \cite{abbasi2019politex,wei2020model} for a more detailed discussion of RL with infinite-time horizon and average reward.

\paragraph{Finite Time Horizon.} When there is a finite time horizon $T<\infty$ we no longer discount future values and have a terminal reward. The MDP problem with finite time horizon can be expressed as
\begin{eqnarray} \label{equ: singlev_fh}
V^\star_t(s) = \sup_{\Pi} V_t^{\Pi}(s):=
\sup_{\Pi}\E^{\Pi}\biggl[\sum_{u=t}^{T-1} r_u(s_u, a_u) +r_T(s_T)\biggl| s_t = s\biggl], \;\; \forall s\in \Sc, 
\end{eqnarray}
subject to
\begin{eqnarray} \label{equ: singledynamics_fh}
s_{u + 1} \sim P_u(s_u, a_u), \;\;\; a_u \sim \pi_u(s_u),\;\; t \leq u \leq T-1.
\end{eqnarray}
As in the infinite horizon case, we denote by $s_u \in \Sc$ and $a_u \in \Ac$ the state and the action of the agent at time $u$. 
However, where this differs from the infinite horizon case, is that we allow time-dependent transition and reward functions. Now $P_u:  \Sc \times \Ac \to \Pc(\Sc)$ denotes the transition function  and $r_u(s, a)$ a real-valued random variable for each pair $(s, a)\in \Sc \times \Ac$ for $t \leq u \leq T-1$. Similarly $r_T(s)$, the terminal reward, is a real-valued random variable for all $s\in \Sc$. Finally, the Markovian policy $\Pi =\{\pi_u\}_{t=0}^T$ can be either deterministic or randomized.

The corresponding Bellman equation for the value function \eqref{equ: singlev_fh} is defined as
\begin{eqnarray} \label{equ:classicalV_fh}
V^\star_t(s) = \sup_{a \in \Ac} \Big\lbrace \E[r_t(s, a)] + \E_{s' \sim P_{t}(s, a)}[V^\star_{t+1}(s')] \Big\rbrace,
\end{eqnarray}
with the terminal condition $V^\star_T(s) = \mathbb{E}[r_{T}(s)]$.
We write the value function as
$$V^\star_t(s) = \sup_{a \in \Ac} Q^\star_t(s, a),$$ 
where the $Q^\star_t$ function is defined to be
\begin{eqnarray} \label{defsingleQ_fh}
Q^\star_t(s, a) = \E[r_t(s, a)] + \E_{s' \sim P_t(s, a)}[V^\star_t(s')].
\end{eqnarray}
There is also a Bellman equation for the $Q$-function given by
\begin{eqnarray} \label{equ: singleQ_fh}
Q^\star_t(s, a) = \E[r_t(s, a)] + \E_{s' \sim P_t(s, a)}\left[\sup_{a' \in \Ac} Q^\star_{t+1}(s', a')\right],
\end{eqnarray}
with the terminal condition $Q^\star_T(s,a) = \mathbb{E}[r_T(s)]$ for all $a\in \mathcal{A}$. We note that since financial time series data are typically non-stationary, time-varying transition kernels and reward functions in \eqref{equ: singlev_fh}-\eqref{equ: singledynamics_fh} are particularly important for financial applications.

For an infinite horizon MDP with finite state and action space, and  finite reward $r$, a further useful observation is that such an MDP always has a stationary optimal policy, whenever an optimal policy exists. 
\begin{Theorem}[Theorem 6.2.7 in \cite{puterman2014markov}]\label{thm:MDP_discounted}Assume $|\mathcal{A}|<\infty$, $|\mathcal{S}|<\infty$ and $|r|<\infty$ with probability one.
 For any infinite horizon discounted MDP, there always exists a deterministic stationary policy that is optimal.
\end{Theorem}

Theorem \ref{thm:MDP_discounted} implies that there always exists a fixed policy so that taking actions specified by that policy at each time step maximizes the discounted reward. The agent does not need to change policies with time. There is a similar result for the average reward case, see Theorem 8.1.2 in \cite{puterman2014markov}. This insight reduces the question of finding the best sequential decision-making strategy to the question of finding the {\it best stationary policy}. To this end, we write $\pi:\mathcal{S}\rightarrow \mathcal{P}(\mathcal{A})$ (without a time index) for a stationary policy throughout the paper.

\paragraph{Linear MDPs and Linear Functional Approximation.}  Linear MDPs have been extensively studied  in the recent literature on theoretical RL in order to establish tractable and efficient algorithms. In a linear MDP, both the transition kernels and the reward function are assumed to be linear with respect to some feature mappings \cite{bradtke1996linear,melo2007q}. 

In the infinite horizon setting, an MDP is said to be linear with a feature map $\phi: \mathcal{S}\times \mathcal{A} \rightarrow \mathbb{R}^d$, if  there exist $d$ unknown (signed) measures $\mu = (\mu^{(1)},\cdots,\mu^{(d)})$ over $\mathcal{S}$ and an unknown vector $\theta \in \mathbb{R}^d$ such that for any $(s,a)\in \mathcal{S}\times \mathcal{A}$, we have
\begin{eqnarray}\label{eq:linearMDP_infinite}
P(\cdot|s,a) = \langle\, \phi(s,a),\mu(\cdot)\,\rangle,\quad r(s,a) = \langle\, \phi(s,a),\theta\,\rangle.
\end{eqnarray}

Similarly for the finite horizon setting, an  MDP is said to be linear with a feature map $\phi: \mathcal{S}\times \mathcal{A} \rightarrow \mathbb{R}^d$, if for any $0\leq t \leq T$, there exist $d$ unknown (signed) measures $\mu_t = (\mu_t^{(1)},\cdots,\mu_t^{(d)})$ over $\mathcal{S}$ and an unknown vector $\theta_t \in \mathbb{R}^d$ such that for any $(s,a)\in \mathcal{S}\times \mathcal{A}$, we have
\begin{eqnarray}\label{eq:linearMDP_finite}
P_t(\cdot|s,a) = \langle\, \phi(s,a),\mu_t(\cdot)\,\rangle,\quad r_t(s,a) = \langle\, \phi(s,a),\theta_t\,\rangle,
\end{eqnarray}
Typically features are assumed to be known to the agent and bounded, namely, $\|\phi(s,a)\|\leq 1$ for all $(s,a)\in \mathcal{S}\times \mathcal{A}$.

The linear MDP framework is closely related to the literature on RL with {\it linear functional approximation}, where the value function is assumed to be of the form
\begin{eqnarray}\label{eq:lfa_infinite}
Q(s,a) = \langle\, \psi(s,a), \omega \,\rangle, \quad V(s) = \langle\, \xi(s), \eta \,\rangle
\end{eqnarray}
for the infinite horizon case, and
\begin{eqnarray}\label{eq:lfa_finite}
Q_t(s,a) = \langle\, \psi(s,a), \omega_t \,\rangle, \quad V_t(s) = \langle\, \xi(s), \eta_t \,\rangle, \forall\, 0\leq t \leq T
\end{eqnarray}
for the finite horizon case. Here $\psi: \mathcal{S}\times \mathcal{A}\rightarrow \mathbb{R}^d$ and $\xi: \mathcal{S}\rightarrow \mathbb{R}^d$ are known feature mappings and $\omega$, $\omega_t$ $\eta$, and $\eta_t$ are unknown vectors. It has been shown that the linear MDP (i.e., \eqref{eq:linearMDP_infinite} or \eqref{eq:linearMDP_finite}) and the linear functional approximation formulation (i.e., \eqref{eq:lfa_infinite} or \eqref{eq:lfa_finite}) are equivalent under mild conditions \cite{jin2020provably,yang2019sample}. In addition to linear functional approximations, we refer the reader to \cite{dai2018sbeed} for a general discussion of RL with nonlinear functional approximations and to Section \ref{sec:deep_value_based} for neural network approximation,  one of the most popular nonlinear functional approximations used in practice.

\paragraph{Nonlinear Functional Approximation.} Compared to linear functional approximations, nonlinear functional approximations do not require knowledge of the kernel functions a priori. Nonlinear functional approximation could potentially address the mis-specification issue when the agent has an incorrect understanding of the functional space that the MDP lies in. The most popular nonlinear functional approximation approach is to use neural networks, leading to deep RL. Thanks to the universal approximation theorem, this is a theoretically sound approach and neural networks show promising performance for 
a wide range of applications. Meanwhile gradient-based algorithms for certain neural network architectures enjoy provable convergence guarantees. We defer the discussion of deep RL to Section \ref{sec:deep_value_based} and its financial applications to Section \ref{sec:fin_app}.

\subsection{From MDP to Learning}\label{sec:MDP2learning}
When the transition dynamics $P$ and the reward function $r$ for the infinite horizon MDP problem are unknown, this MDP becomes a reinforcement learning problem, which is to find an optimal stationary policy $\pi$ (if it exists) 
while simultaneously learning the unknown $P$ and $r$. The learning of $P$ and $r$ can be either explicit or implicit, which leads to model-based and model-free RL, respectively. The analogous ideas hold for the finite horizon case. We introduce some standard RL terminology. A more detailed introduction to RL can be found in textbooks such as \cite{suttonbarto,powell2021reinforcement}.

\paragraph{Agent-environment Interface.} In the RL setting, the learner or the decision-maker is called the agent. The physical world that the agent operates in and interacts with, comprising
everything outside the agent, is called the environment. The agent and environment interact at each of a sequence of discrete time steps, $t =0,1,2,3,\cdots$, in the following way. At the beginning of each time step $t$, the agent receives some representation of the environment's state, $s_t \in \mathcal{S}$ and selects an action $a_t\in \mathcal{A}$. At the end of this time step, in part as a consequence of its action, the agent receives a numerical reward $r_t$ (possibly stochastic) and a new state $s_{t+1}$ from the environment. The tuple $(s_t,a_t,r_t,s_{t+1})$ is called a sample at time $t$ and $h_t :=\{(s_u,a_u,r_u,s_{u+1})\}_{u=0}^t$ is referred to as the history or experience up to time $t$. An RL algorithm is a finite sequence of well-defined policies for the agent to interact with the environment.
 
\paragraph{Exploration vs Exploitation.}  In order to maximize the accumulated reward over time, the agent
learns to select her actions based on her past experiences (exploitation) and/or by trying new choices (exploration).  Exploration provides opportunities to improve performance from the current sub-optimal solution to the ultimate globally optimal one, yet it is time consuming and computationally expensive as over-exploration may impair the convergence to the optimal solution. Meanwhile, pure exploitation, i.e., myopically picking the current solution based solely on
past experience, though easy to implement, tends to yield sub-optimal global solutions. Therefore,
an appropriate trade-off between exploration and exploitation is crucial in the design of RL algorithms in order to improve the learning and the optimization performance. See Section \ref{sec:ex-ex} for a more detailed discussion.

\paragraph{Simulator.} In the procedure described above on how the agent interacts with the environment, the agent does so in an online fashion. Namely, 
the initial state at time step $t+1$, $s_{t+1}$, is the state that the agent moves to after taking action $a_t$ from state $s_t$. This is a challenging setting since an efficient exploration scheme is needed. For instance, $Q$-learning with the $\varepsilon$-greedy policy (to be introduced in Section \ref{sec:sarsa}) may take exponentially many episodes
to learn the optimal policy \cite{kearns2002near}.

Some results assume access to a simulator \cite{koenig1993complexity} (a.k.a., a generative model \cite{azar2012sample}),
which allows the algorithm to query {\it arbitrary} state-action pairs and return the reward and the next state. This implies the agent can ``restart'' the system at any time. That is, the initial state at time step $t+1$  does not need to be the state that agent moves to after taking action $a_t$ from state $s_t$. The ``simulator'' significantly alleviates the difficulty of exploration, since a naive exploration strategy which queries all state-action pairs uniformly at random already leads to the most efficient algorithm for finding an optimal policy \cite{azar2012sample}.  

\paragraph{Randomized Policy vs Deterministic Policy.} 
A randomized policy $\pi_t:\Sc \rightarrow \mathcal{P}(\Ac)$ is also known in the control literature as a relaxed control and in game theory as a mixed strategy. Despite the existence of a {\it deterministic} optimal policy as stated in Theorem  \ref{thm:MDP_discounted} for the infinite time horizon case, most of the RL algorithms adopt randomized policies to encourage exploration when RL agents are not certain about the environment.

\paragraph{An Example: Multi-armed bandits.} We illustrate these points by discussing
multi-armed bandit problems, a special case of RL problems. 
The multi-armed bandit is a model for a set of slot machines. A simple version is that there are a number of arms, each with a stochastic reward coming from a fixed probability distribution which is initially unknown.
We try these arms in some order, which may depend on the sequence of rewards that have been
observed so far and attempt to maximize our return. A common objective in this context is to find a policy for choosing the next arm
to be tried, under which the sum of the expected rewards comes as close as possible to the ideal
reward, i.e., the expected reward that would be obtained if we were to try the ``best'' arm at all times. 
Thus the agent interacts with the environment by selecting an action - pulling an arm - and receiving a reward.  
The policy of the order of pulling on the arms has to balance exploitation, the continued use of an arm that is returning a decent reward, with exploration, sampling arms about which there is less information to find one with a possibly better return.

In this setting for a multi-armed bandit problem we have not involved state dynamics. In this case, an admissible policy is simply a distribution over the action space for a randomized policy and a single arm for a deterministic policy. A simple example of a bandit with a state space is the stochastic contextual bandit which is used extensively in the personalized advertisement recommendation and clinical treatment recommendation literature. In this case the state dynamics (also referred to as the context) are sampled i.i.d. from an unknown distribution. For the personalized advertisement recommendation example, the state at time $t$ can be interpreted as the personal information and purchasing behavior of a targeted client. 

The problem was first considered in the seminal work of Robbins \cite{robbins1952some}, which 
derives policies that asymptotically attain an average reward that converges in the limit to the reward of the best arm.
The multi-armed bandit problem was later studied in discounted, Bayesian, Markovian, expected
reward, and adversarial setups. See Berry and Fristedt \cite{berry1985bandit} for a review of the classical results on the multi-armed bandit problem. 


\subsubsection{Performance Evaluation}

It is not possible to solve MDPs analytically in general and therefore we will discuss numerical algorithms to determine the optimal policies. 
In order to assess different algorithms we will need a measure of their performance. Here we will consider several types of performance measure: sample complexity, rate of convergence, regret analysis and asymptotic convergence.

For RL with a finite time horizon, one episode contains a sequence of states, actions and rewards, which starts at time $0$ and ends at the terminal time $T$, and the performance is evaluated in terms of the total number of samples in all episodes.\footnote[1]{Sample complexity with respect to the number of episodes has also been adopted in the literature \cite{DannBrunskill2015,DannUBEV}. The translation between these two criteria is straightforward.} Sometimes we also refer to this setting as episodic RL. For RL with infinite time horizon, the performance is evaluated in terms of the number of steps. In this setting one step contains one (state, action, reward, next state) tuple. It is worth noting that the episodic criterion can also be used to analyze infinite horizon problems. One popular technique is to truncate the trajectory (with infinite steps) at a large time and hence translate the problem into an approximating finite time horizon problem. Examples include REINFORCE \cite{zhang2020sample} and the policy gradient method \cite{liu2020improved} (see Section~\ref{subsec:policy_based_methods}). 

Throughout the analysis, we use the notation $\widetilde{O}(\cdot)$  to hide logarithmic factors when describing the order of the scaling in $\varepsilon$, $\delta$, $|\mathcal{A}|$ and $|\mathcal{S}|$ (when these are finite), and other possible model parameters such as the discount rate $\gamma$ in the infinite horizon case or the dimension $d$ of the features when functional approximation is used. We write $\widetilde{\mathcal{O}}^{\prime}$ rather than $\widetilde{\mathcal{O}}$ to indicate that, although there may be other parameters, we only include the dependence on $\varepsilon$. 
We denote by $\mbox{poly}(\cdot)$ a polynomial function of its arguments.

\paragraph{Sample Complexity.} In RL, a sample $(s,a,r,s^{\prime})$ is defined as a tuple consisting of a state $s$, an action $a$, an instantaneous reward received by taking action $a$ at state $s$, and the next state $s^{\prime}$ observed afterwards. Sample complexity is defined as the total number of samples required to find an approximately optimal policy. This evaluation criterion can be used for any kind of RL problem.

For episodic RL, an algorithm returns, after the end of the $(k-1)$-th episode with $M_{k-1}$ samples used in this episode, a policy $\pi_k$ to be applied in the $k$-th episode.
The sample complexity of this algorithm is the minimum number $\sum_{k=1}^K M_{k-1}(\varepsilon)$ of samples such that for all $k \ge K$, $\pi_k$ is $\varepsilon$-optimal with probability at least $1-\delta$, i.e., for $k \ge K$,
\begin{eqnarray}\label{eq:sample_complexity_epi}
\mathbb{P}\left(V_0^{\pi_k}\ge V_0^\star-\varepsilon\right) \ge 1-\delta.
\end{eqnarray}

For discounted RL, an algorithm returns, after
the end of the $(n-1)$-th step with $M_{n-1}$ samples used in this iteration, a policy $\pi_n$ to be applied in the $n$-th step. There are several notions of sample complexity in this setting.
The most commonly used ones are defined through the value function and the $Q$-function with all input variables (i.e., $s\in \mathcal{S}$ for $V$ and $(s,a)\in \mathcal{S}\times \mathcal{A}$ for $Q$) and they are referred to as the $V$-sample complexity and $Q$-sample complexity in the literature.  The $Q$-sample complexity of a given algorithm is defined as the minimum number $\sum_{n=1}^N M_{n-1}(\varepsilon)$ of samples such that for all $n \ge N$, $Q^{(n)}$ is $\varepsilon$-close to $Q^*$, that is
\begin{eqnarray}\label{eq:sample_complexity_infi_Q}
\|Q^{(n)} - Q^*\|_{\infty}:=\sup_{s\in\mathcal{S},a\in \mathcal{A}}\Big|Q^*(s,a)-Q^{(n)}(s,a)\Big| \leq \varepsilon,
\end{eqnarray}
holds either with high probability (that is $\mathbb{P}\left(\|Q^{(n)}-Q^*\|_{\infty}<\varepsilon\right) \ge 1-\delta$) or in the expectation sense $\mathbb{E}[\|Q^{(n)}-Q^*\|_{\infty}]<\varepsilon$.
Similarly, the $V$-sample complexity is defined as the minimum number $\sum_{n=1}^N M_{n-1}(\varepsilon)$ of samples such that for all $n \ge N$, $V^{(n)}$ is $\varepsilon$-close to $V^*$, that is
\begin{eqnarray}\label{eq:sample_complexity_infi_V_2}
\|V^{(n)} - V^*\|_{\infty}:=\sup_{s\in\mathcal{S}}\left|V^{(n)}(s)-V^*(s)\right|\leq \varepsilon,
\end{eqnarray}
holds either with high probability (in that $\mathbb{P}\left(\|V^{(n)}-V^*\|_{\infty}<\varepsilon\right) \ge 1-\delta$) or in the expectation sense $\mathbb{E}[\|V^{(n)}-V^*\|_{\infty}]<\varepsilon$. 
Note that \eqref{eq:sample_complexity_infi_Q} implies \eqref{eq:sample_complexity_infi_V_2} since $V^{(n)}(s) = \sup_{a\in \mathcal{A}}Q^{(n)}(s,a)$ and $V^{*}(s) = \sup_{a\in \mathcal{A}}Q^*(s,a)$. 
In our analysis we do not distinguish between whether the sample complexity bounds for \eqref{eq:sample_complexity_infi_Q} and \eqref{eq:sample_complexity_infi_V_2} are in the high probability sense or in the expectation sense.

The second type of sample complexity, the {\it sample complexity of exploration}, is  defined as the number of samples such that the non-stationary policy $\pi_n$ at time $n$ is not $\varepsilon$-optimal for the current state $s_n$. This measure counts the number of mistakes along the whole trajectory. Mathematically speaking, the sample complexity of exploration for a given algorithm is the minimum number $\sum_{n=1}^N M_{n-1}(\varepsilon)$ of samples such that for all $n \ge N$, $\pi_n$ is $\varepsilon$-optimal when starting from the current state $s_n$ with probability at least $1-\delta$, i.e., for $n \ge N$,
\begin{eqnarray}\label{eq:sample_complexity_infi_V}
\mathbb{P}\left(\left|V^{\pi_n}(s_n)- V^\star(s_n)\right|\leq \varepsilon\right) \ge 1-\delta.
\end{eqnarray}
Note that the condition $\mathbb{P}\left(\|V^{(n)}-V^*\|_{\infty}<\varepsilon\right) \ge 1-\delta$ associated with \eqref{eq:sample_complexity_infi_V_2} is stronger and implies the condition in \eqref{eq:sample_complexity_infi_V}. 

Another weaker criterion, the {\it mean-square sample complexity}, measures the sample complexity in the average sense with respect to the steady state distribution or the initial distribution $\mu$.  
In this case, the  {\it mean-square sample complexity} is defined as the minimum number $\sum_{n=1}^N M_{n-1}(\varepsilon)$ of samples such that for all $n \ge N$,
\begin{eqnarray}\label{eq:sample_complexity_infi_V_avg}
\|V^{\pi_n} - V^*\|_{\mu}:=\sqrt{\int_{\mathcal{S}}(V^{\pi_n}(s)-V^*(s))^2\mu(ds)} \leq \varepsilon.
\end{eqnarray}
Since \eqref{eq:sample_complexity_infi_V_avg} is an aggregated guarantee via the $V$ function over a state distribution $\mu$, it is implied by \eqref{eq:sample_complexity_infi_V_2}.



\paragraph{Rate of Convergence.} To emphasize the convergence with respect to the number of iterations required, we introduce the notion of \emph{rate of convergence}, which uses the relationship between the number of iterations/steps $N$ and the error term $\varepsilon$, to quantify how fast the learned policy converges to the optimal solution. This is also called the iteration complexity in the RL and optimization literature. Compared to the notion of sample complexity introduced above, the rate of convergence calculates the number of iterations needed to reach a given accuracy while ignoring the number of samples used within each iteration. When only one sample is used in each iteration, the rate of convergence coincides with the sample complexity. In addition when a constant number of samples (i.e., independent from $\varepsilon$) are used within each iteration, the rate of convergence and the sample complexity are of the same order with respect to  $\varepsilon$.  

In particular, an algorithm is said to converge at a \emph{linear} rate if $N\sim \widetilde{\mathcal{O}}^{\prime}(\log(1/\varepsilon))$. Similarly, an algorithm is said to converge at a \emph{sublinear} rate (slower than linear) if $N\sim \widetilde{\mathcal{O}}^{\prime}(1/\varepsilon^p)$ with $p\geq 1$, and is said to converge at a \emph{superlinear} (faster than linear) if $N\sim \widetilde{\mathcal{O}}^{\prime}(\log(\log(1/\varepsilon))$. 

\paragraph{Regret Analysis.} The {\it regret} of a given policy $\pi$ is defined as the difference between the cumulative reward of the optimal policy and that gathered by $\pi$. It quantifies the exploration-exploitation trade-off.

For episodic RL with horizon $T$ in formulation \eqref{equ: singlev_fh}, the regret of an algorithm after $K $ episodes with $M=T\,K$  steps is
\begin{eqnarray}\label{eq:regret_epi}
R(M) = \left\| K \times V_0^{*} - \sum_{k=1}^K\mathbb{E}[V_0^{\pi_k}]\right\|_{\infty} =  \sup_{s\in \mathcal{S}} \Big(K \times V_0^{*} - \sum_{k=1}^K\mathbb{E}[V_0^{\pi_k}]\Big).
\end{eqnarray}
There is currently no regret analysis for RL problems with infinite time horizon and discounted reward. The natural definition of regret as given above is less meaningful in this case as the cumulative discounted reward is bounded and hence does not scale with $T$. 

\paragraph{Asymptotic Convergence.} In addition to sample complexity and regret analysis discussed above, \emph{asymptotic} convergence is another weaker convergence guarantee, which requires the error term to decay to zero as $N$ goes to infinity (without specifying the order of $N$). This is often a first step in the theoretical analysis of RL algorithms.


\subsubsection{Classification of RL Algorithms}


An RL  algorithm includes one or more of the following components:
\begin{itemize}
    \item a representation of a value function that provides a prediction of how good each state or each state/action pair is,
    \item  a direct representation of the policy $\pi(s)$ or $\pi(s,a)$, 
\item  a model of the environment (the estimated transition function and the estimated reward function) in conjunction with a planning algorithm (any computational process that uses a model to create or improve a policy).
\end{itemize}
The first two components are related to what is called model-free RL. When the latter component is used, the algorithm is referred to as model-based RL.

In the MDP setting, model-based algorithms maintain an approximate MDP model by estimating the transition probabilities and the reward function, and derive a value function from the approximate MDP. The policy is then derived from the value function. Examples include \cite{brafman2002r,kearns2002near,lattimore2012pac} and \cite{MoRmax2010}. Another line of model-based algorithms make structural assumptions on the model, using some prior knowledge, and utilize the structural information for algorithm design. For example, see \cite{fazel2018global} for learning linear-quadratic regulators over an infinite time horizon  and \cite{hambly2020policy,basei2021logarithmic} for the finite time horizon case.

Unlike the model-based method, model-free algorithms {\it directly} learn a value (or state-value) function or the optimal policy without inferring the model.
Model-free algorithms can be further divided into two categories, value-based methods and policy-based methods. Policy-based methods explicitly build a representation of a policy and keep it in memory during learning. Examples include policy gradient methods and trust-region policy optimization methods. As an alternative,  value-based methods store only a value function without an explicit policy during the learning process. In this case, the policy is implicit and can be derived directly from the value function (by picking the action with the best value). 

In our discussion of methodology, we focus on model-free RL algorithms for MDP with infinite horizon and discounted reward. In particular,  we introduce some classical value-based and policy-based methods in Sections \ref{sec:value_based_methods} and \ref{subsec:policy_based_methods} respectively. For the episodic setting and model-based algorithms, see the discussion in Section \ref{sec:discussion}.

\subsection{Value-based Methods}\label{sec:value_based_methods}
In this section, we introduce the methodologies of several classical value-based methods in the setting of an infinite time horizon with discounting, finite state and action spaces ($|\mathcal{A}|<\infty$ and $|\mathcal{S}|<\infty$), and stationary policies. 
The setting with finite state and action spaces is also referred to as the tabular case in the RL literature. For more general setups with an infinite number of actions and states, we refer the reader to Section \ref{sec:value_discussion} for a discussion of linear functional approximations and to Section \ref{subsec:deep_value_based} for neural network approximations.

Recall that for reinforcement learning, the distribution of the reward function $r$ and the transition function $P$ are unknown to the agent. Therefore the expectations in the Bellman equations \eqref{equ:classicalV} and \eqref{equ: singleQ} cannot be calculated directly. On the other hand, the agents can observe samples $(s,a,r,s^{\prime})$ by interacting with the system without a model of the system's dynamics $P$ or any prior knowledge of $r$. The agents can then learn the value function or $Q$-function using the samples. Following this idea, we next introduce the classical temporal-difference learning algorithm, with $Q$-learning and State-Action-Reward-State-Action (SARSA) as two special cases.

\subsubsection{Temporal-Difference Learning}
Given some samples $(s,a,r,s')$ obtained by following a policy $\pi$, the agent can update her estimate of the value function $V^{\pi}$ (defined in \eqref{equ: singlev}) at the $(n+1)$-th iteration by
\begin{eqnarray}\label{eq:TD_update}
V^{\pi,(n+1)}(s) \leftarrow (1-\beta_n(s,a))\underbrace{V^{\pi,(n)}(s)}_{\text{current estimate}} + \beta_n(s,a) \underbrace{[r+\gamma V^{\pi,(n)}(s')]}_{\text{new estimate}},
\end{eqnarray}
with some initialization of the value function $V^{\pi,(0)}$. Here
$\beta_n(s,a)$ is the learning rate at iteration $n+1$ which balances the weighting between the current estimate and the new estimate.  The quantity $\beta_n$ can be a constant, can depend on the current state $s$, or even depend on the observed samples up to iteration $n+1$. Algorithm \ref{alg:TD} provides some pseudo-code for an implementation of the TD method. Equation
\eqref{eq:TD_update} is sometimes referred to as a TD(0) method. This is a special case of a TD$(\lambda)$ method (for some $\lambda \in [0,1]$) which is a combination of TD learning (with weight $\lambda$) and a Monte Carlo method (with weight $1-\lambda$). Here a Monte Carlo method indicates a simulation-based method to calculate the value function with a longer period of data (in the episode-by-episode sense) rather than a single sample in each update. See Section \ref{sec:REINFORCE} for a detailed discussion of the REINFORCE method which is an example of such a Monte Carlo method.
Equation \eqref{eq:TD_update} is also referred to as a one-step TD method as it only uses information from one update step of the system. There are multi-step TD methods; see Chapter 7 and Chapter 12 in \cite{suttonbarto} for more details.

The TD update in \eqref{eq:TD_update} can also be written as
\begin{eqnarray}\label{eq:TD_update_v2}
V^{\pi,(n+1)}(s) \leftarrow V^{\pi,(n)}(s) + \beta_n(s,a) \underbrace{\left[\overbrace{r+\gamma V^{\pi,(n)}(s')}^{\text{TD target}}-V^{\pi,(n)}(s)\right]}_{\text{TD error $\delta_n$}}.
\end{eqnarray}
The term $\delta_n$, commonly called the TD error, measures the difference between the estimated value at $s$ and the better estimate $r+\gamma V^{\pi,(n)}(s')$, which is often called the  TD target in the literature. The TD error and TD target are two important components in analyzing the convergence of the approximate value function and arise in various forms in the reinforcement learning literature. 

\begin{algorithm}[H]
\caption{\textbf{TD(0) Method for estimating $V^{\pi}$}}
\label{alg:TD}
\begin{algorithmic}[1]
    \STATE \textbf{Input}: total number of iterations $N$; the policy $\pi$ used to sample observations; rule to set learning rate $\beta_n \in (0,1]$ $(0\leq n \leq N-1)$
    \STATE Initialize $V^{\pi,(0)}(s)$ for all $s\in \mathcal{S}$
    \STATE Initialize $s$
    \FOR {$n=0,\cdots,N-1$}
    \STATE Sample action $a$ according to $\pi(s)$
    \STATE Observe $r$ and $s'$ after taking action $a$
   \STATE Update $V^{\pi,(n+1)}$ according to \eqref{eq:TD_update} with $(s,a,r,s')$
   \STATE $s\leftarrow s'$
    \ENDFOR
\end{algorithmic}
\end{algorithm}

\subsubsection{$Q$-learning Algorithm}

A version of TD learning is $Q$-learning. This is a stochastic approximation to the Bellman equation \eqref{equ: singleQ} for the $Q$-function with samples observed by the agent. At iteration $n$ ($1 \leq n \leq N-1$), the $Q$-function is updated using a sample $(s,a,r,s^{\prime})$ where $s' \sim P(s,a)$,
\begin{eqnarray}\label{eq:Q_learning_update}
Q^{(n+1)}(s,a)\leftarrow (1-\beta_n(s,a))\underbrace{Q^{(n)}(s,a)}_{\text{current estimate}}+\beta_n(s,a)\underbrace{\left[r(s,a)+\gamma\max_{a'}Q^{(n)}(s',a')\right]}_{\text{new estimate}}.
\end{eqnarray}
Here $\beta_n(s,a)$ is the learning rate which balances the weight between the current estimate $Q^{(n)}$ from iteration $n$ and the new estimate $r(s,a)+\gamma\max_{a'}Q^{(n)}(s',a')$ calculated using the sample $(s,a,r,s^{\prime})$. Algorithm \ref{alg:Q} provides pseudo-code for an implementation of  Q-learning.

\begin{algorithm}[H]
\caption{\textbf{$Q$-learning with samples from a policy $\pi$}}
\label{alg:Q}
\begin{algorithmic}[1]
    \STATE \textbf{Input}: total number of iterations $N$; the policy $\pi$ to be evaluated; rule to set learning rate $\beta_n \in (0,1]$ $(0 \leq n\leq N-1)$
    \STATE Initialize $Q^{(0)}(s,a)$ for all $s\in \mathcal{S}$ and $a\in \mathcal{A}$
    \STATE Initialize $s$
    \FOR {$n=0,\cdots,N-1$}
    \STATE Sample action $a$ according to $\pi(s)$
    \STATE Observe $r$ and $s'$ after taking action $a$
   \STATE Update $Q^{(n+1)}$ according to \eqref{eq:Q_learning_update} with sample $(s,a,r,s')$
   \STATE $s\leftarrow s'$
    \ENDFOR
\end{algorithmic}
\end{algorithm}

The following theorem guarantees the asymptotic convergence of the $Q$-learning update \eqref{eq:Q_learning_update} to the $Q$-function defined in \eqref{defsingleQ}.
\begin{Theorem}[Watkins and Dayan (1992) \cite{watkins1992q}] \label{thm:Q_conv} Assume $|\mathcal{A}|<\infty$ and $|\mathcal{S}|<\infty$.
Let $n^i(s, a)$ be the index of the $i$-th time that the action $a$ is used in state $s$.
Let $R<\infty$ be a constant. Given bounded rewards $|r|\leq R$, learning rate $0\leq\beta_n<1$ and 
\begin{eqnarray}\label{exploration}
\sum_{i=1}^{\infty}\beta_{n^i(s,a)} = \infty,\,\,\sum_{i=1}^{\infty}(\beta_{n^i(s,a)})^2  < \infty, \,\,\forall s,a,
\end{eqnarray}
then $Q^{(n)}(s,a)\rightarrow Q^\star(s,a)$ as $n\rightarrow \infty$, $\forall s,a$ with probability $1$.
\end{Theorem}

This is a classical result (one of the first few theoretical results in RL) and the proof of the convergence is based on  the action-replay process (ARP), which is an artificial controlled Markov process constructed from the episode sequence and the learning rate sequence $\beta$. Theorem \ref{thm:Q_conv} implies that eventually $Q^{(n)}$ will convergence to the true value function $Q^\star$ when condition \eqref{exploration} is satisfied. However, this asymptotic result does not provide any insights on how many samples are needed to reach a given accuracy. More results on the sample complexity for $Q$-learning related algorithms are discussed in Section~\ref{subsubsec:value_theo}.

\subsubsection{SARSA}\label{sec:sarsa}
In contrast to the $Q$-learning algorithm, which takes samples from any policy $\pi$ as the input where these samples could be collected in advance before performing the $Q$-learning algorithm, SARSA adopts a policy which is based on the agent's current estimate of the $Q$-function. The different source of samples is indeed the key difference between on-policy and off-policy learning, which will be discussed after the SARSA algorithm. 

\begin{algorithm}[H]
\caption{\textbf{SARSA: On-policy TD Learning}}
\label{alg:SARSA}
\begin{algorithmic}[1]
    \STATE \textbf{Input}: total number of iterations $N$, the learning rate $\beta\in(0,1)$\footnotemark[1], and small parameter $\varepsilon>0$.
    \STATE Initialize $Q^{(0)}(s, a)$ for all $s\in \mathcal{S}$ and $a\in \mathcal{A}$
    \STATE Initialize $s\in \mathcal{S}$
    \STATE Initialize  $a$: Choose $a$ when in $s$ using a policy derived from $Q^{(0)}$ (e.g, $\varepsilon$-greedy)
    \FOR{$n=0,1,\cdots,N-1$}
    \STATE Take action $a$, observe reward $r$ and the next step $s'$
    \STATE Choose $a'$ when in $s'$ using a policy derived from $Q^{(n)}$ (e.g., $\varepsilon$-greedy) 
    \begin{equation}\label{eq:SARSA_update}
    Q^{(n+1)}(s,a) \leftarrow (1-\beta)\underbrace{ Q^{(n)}(s,a)}_{\text{current estimate}} +\beta\underbrace{(r+\gamma Q^{(n)}(s',a'))}_{\text{new estimate}}
    \end{equation}
    \STATE $s\leftarrow s'$, $a\leftarrow a'$
    \ENDFOR 
\end{algorithmic}
\end{algorithm}
\footnotetext[1]{The learning rate can depend on the number of iterations, the state-action pair and other parameters in the algorithm. For notational simplicity, we demonstrate the SARSA Algorithm (and the rest of the algorithms mentioned) with a constant learning rate $\beta$. In the convergence analysis, we use $\beta_n$ to denote a learning rate which depends on the number of iterations.}

The policy derived from $Q^{(n)}$ using $\varepsilon$-greedy (line 4 and line 8 in Algorithm \ref{alg:SARSA}) suggests the following choice of action $a$ when in state $s$:
\begin{eqnarray}\label{epsilon_policy}
\begin{cases}
 \text{ select from }\arg\max_{a'} Q^{(n)}(s,a') & \text{with probability } 1-\varepsilon,\\
\text{ uniformly sample from } \mathcal{A} & \text{with probability } \varepsilon.
\end{cases}
\end{eqnarray}

In Algorithm \ref{alg:SARSA}, SARSA uses the $Q^{(n)}$ function with an $\epsilon$-greedy policy to choose $a'$ and to perform exploration. In contrast, Q-learning uses the maximum value of $Q^{(n)}$ over all possible actions for the next step, which is equivalent to setting $\varepsilon=0$ when choosing $a'$.

In addition to the $\varepsilon$-greedy policy, other exploration methods such as the softmax operator (resulting in a Boltzmann policy) \cite{asadi2017alternative,haarnoja2017reinforcement,WZZ2018} can also be applied. See Section \ref{sec:ex-ex} for a more detailed discussion.

\paragraph{On-policy Learning vs. Off-policy Learning.}  An off-policy agent learns the value of the optimal policy independently of the agent's actions. An on-policy agent learns the value of the policy being carried out by the agent including the exploration steps.  $Q$-learning is an off-policy agent as the samples $(s,a,r,s')$ used in updating \eqref{eq:Q_learning_update} may be collected from any policy and may be independent of the agent's current estimate of the $Q$-function. The convergence of the $Q$-function relies on the exploration scheme of the policy $\pi$ (see Theorem \ref{thm:Q_conv}). In contrast to $Q$-learning, SARSA is an on-policy learning algorithm and uses only its own estimation of the $Q$-function to select an action and receive a real-time sample in each iteration. The difference is seen in how the new estimate in the update is calculated. In the update step of $Q$-learning \eqref{eq:Q_learning_update}, we have $\max_{a'} Q^{(n)}(s',a')$ which is the maximum value of the $Q$-function at a given state $s'$. On the other hand, the new estimate in the update of SARSA \eqref{eq:SARSA_update} takes $Q^{(n)}(s',a')$ where $a'$ is selected based on a policy derived from $Q^{(n)}$ (via the $\varepsilon$-greedy policy \eqref{epsilon_policy}).


\subsubsection{Discussion}\label{subsubsec:value_theo}
In this section, we provide a brief overview of sample complexity 
results for model-free and value-based RL with infinite-time horizon and discounted reward. Sample complexity results for the model-based counterpart are deferred to Section \ref{sec:discussion}.

There has been very little non-asymptotic analysis of TD(0) learning. Existing results have focused on linear function approximations. For example, \cite{dalal2018finite} showed that the mean-squared error converges at a rate of $\widetilde{\mathcal{O}}^{\prime}(\frac{1}{\varepsilon^{1/\sigma}})$ with decaying learning rate $\beta_n= \frac{1}{(1+n)^{\sigma}}$ for some $\sigma\in (0,1)$ where i.i.d. samples are drawn from a stationary distribution; \cite{lakshminarayanan2018linear} provided
an improved $\widetilde{\mathcal{O}}^{\prime}(\frac{1}{\varepsilon})$ bound for iterate-averaged TD(0) with constant step-size; and \cite{bhandari2018finite} reached the same mean-square sample complexity $\widetilde{\mathcal{O}}^{\prime}(\frac{1}{\varepsilon})$ with a simpler proof and extends the framework to the case where non-i.i.d. but Markov samples are drawn from the online trajectory.
A similar analysis was applied by
\cite{zou2019finite} to SARSA and $Q$-learning algorithms for a continuous state space using linear function approximation (see the end of Section \ref{sec:set_up}).

The $Q$-sample complexity for the $Q$-learning algorithm of $\widetilde{\mathcal{O}}\left(\frac{|\mathcal{S}||\mathcal{A}|}{(1-\gamma)^5\varepsilon^{5/2}}\mbox{poly}(\log \delta^{-1})\right)$ was first established in \cite{even2003learning} with  decaying learning rate $\beta_n = \frac{1}{(1+n)^{4/5}}$  and access to a simulator.
This was followed by \cite{beck2012error} which showed that the same order of $Q$-sample complexity $\widetilde{\mathcal{O}}\left(\frac{|\mathcal{S}||\mathcal{A}|}{(1-\gamma)^5\varepsilon^{5/2}}\right)$ could be achieved with a constant learning rate $\beta_n \equiv \frac{(1-\gamma)^4\varepsilon^2}{|\mathcal{A}||\mathcal{S}|}$ $(n\leq N)$. Without access to a simulator, delayed $Q$-learning \cite{strehl2009reinforcement} was shown to achieve a sample complexity of exploration of order $\widetilde{\mathcal{O}}\left(\frac{|\mathcal{S}||\mathcal{A}|}{(1-\gamma)^8\varepsilon^{4}}\mbox{poly}(\log \delta^{-1})\right)$  assuming a 
known reward. An improvement to this result to   $\widetilde{\mathcal{O}}\left(\frac{|\mathcal{S}||\mathcal{A}|}{(1-\gamma)^7\varepsilon^{4}}\mbox{poly}(\log \delta^{-1})\right)$ has recently been obtained using an Upper-Confidence-Bound (UCB) exploration policy \cite{dong2019q} and a single trajectory.  
In addition, sample complexities of $Q$-learning variants with (linear) function approximations have been established in several recent papers.  \cite{yang2019sample} assumed a linear MDP framework (see the end of Section \ref{sec:set_up})
and the authors provided a V-sample complexity $\widetilde{\mathcal{O}}\left(\frac{d}{(1-\gamma)^3\varepsilon^{2}}\mbox{poly}(\log \delta^{-1})\right)$ with $d$ denoting the dimension of the features. This result matches the information-theoretic lower bound up to $\log(\cdot)$ factors. Later work, \cite{lattimore2020learning} considered the setting where the transition kernel can be approximated by linear functions with small approximation errors to make the model more robust to model mis-specification. In this case, the authors provided a V-sample complexity  of order $\widetilde{\mathcal{O}}\left(\frac{K}{(1-\gamma)^4\varepsilon^{2}}\mbox{poly}(\log \delta^{-1})\right)$. Although the dependence on $\gamma$ is not as good as in \cite{yang2019sample}, this framework can be applied to a more general class of models which are ``near'' linear. Although $Q$-learning is one of the most successful algorithms for finding the best
action-value function $Q$, its implementation often suffers from large high biased estimates of the $Q$-function values incurred by random sampling. The double $Q$-learning algorithm proposed in \cite{hasselt2010double}
tackled this high estimation issue by randomly switching the update between two $Q$-estimators, and has thus gained significant popularity in practice. The first sample complexity result for the double $Q$-learning algorithm was established in \cite{xiong2020finite} with a Q-sample complexity on the order of $\widetilde{\mathcal{O}}\left(\left(\frac{1}{(1-\gamma)^6\varepsilon^2}\ln\,\frac{|\mathcal{A}||\mathcal{S}|}{(1-\gamma)^7\varepsilon^2\delta}\right)^{1/\omega}+\left(\frac{1}{1-\gamma}\ln\,\frac{1}{(1-\gamma)^2\varepsilon}\right)^{1/(1-\omega)}\right)$ and learning rate $\beta_n = \frac{1}{n^{w}}$ for some constant $\omega \in (0,1)$.

\subsection{Policy-based Methods}\label{subsec:policy_based_methods}


 So far we have focused on value-based methods where we learn the value function to generate the policy using, for instance, the  $\varepsilon$-greedy approach. However, these methods may suffer from high computational complexity as the dimensionality of the state or action space increases (for example in continuous or unbounded spaces). In this case, it may be more effective to directly parametrize the policy rather than the value function. Furthermore, it is empirically observed that policy-based methods have better convergence properties than value-based methods \cite{sewak2019policy,yu2020policy,daberius2019deep}. This stems from the fact that value-based methods  can have big oscillations during the training process since the choice of actions may need to change dramatically in order to have an arbitrarily small change in the estimated value functions.

In this section, we focus on model-free policy-based methods, where we learn a parametrized policy without inferring the value function in the setting of infinite time horizon with discounting, finite state and action spaces, and stationary policies. Examples of policy-based methods include REINFORCE \cite{williams1992}, Actor-Critic methods \cite{konda2000}, Trust Region Policy Optimization (TRPO) \cite{schulman2015}, Proximal Policy Optimization (PPO) \cite{schulman2017proximal} among others. We first parametrize the policy $\pi$ by a vector $\theta$, thus we define  $\pi(s,\cdot;\theta)\in \mathcal{P}(\mathcal{A})$,  the probability distribution parameterized by $\theta$ over the action space given the state $s$ at time $t$. Here we will use $\pi(s,a;\theta)$ and $\pi_{\theta}$ interchangeably to represent the policy parameterized by $\theta$. We write $\mu^{\pi_{\theta}}$ for the stationary distribution of the state under policy $\pi(s,a;\theta)$. Then the policy objective function is defined to be
\[
J(\theta):=\int_{\mathcal{S}}V^{\pi_{\theta}}(s)\mu^{\pi_{\theta}}(ds)
\]
for RL with infinite time horizon, or 
\[
J(\theta):=\int_{\mathcal{S}}V_0^{\pi_{\theta}}(s)\mu^{\pi_{\theta}}(ds)
\]
for RL with finite time horizon. In order to maximize this function, the policy parameter $\theta$ is updated using the gradient ascent rule:
\begin{equation}\label{eqn:grad_ascent}
    \theta^\prime = \theta + \beta\,\widehat{\nabla_{\theta}J(\theta)},
\end{equation}
where $\beta$ is the learning rate, and $\widehat{\nabla_{\theta}J(\theta)}$ is an estimate of the gradient of $J(\theta)$ with respect to $\theta$.

\subsubsection{Policy Gradient Theorem}

Our task now is to estimate the gradient $\nabla_{\theta}J(\theta)$. The objective function $J(\theta)$ depends on both the policy and the corresponding stationary distribution $\mu^{\pi_{\theta}}$, and the parameter $\theta$ affects both of them. Calculating the gradient of the action with respect to $\theta$ is straightforward given the parametrization $\pi(s,a;\theta)$, whereas it might be challenging to analyze the effect of $\theta$ on the state when the system is unknown. Fortunately the well-known \emph{policy gradient theorem} provides an expression for $\nabla_{\theta}J(\theta)$ which does not involve the derivatives of the state distribution with respect to $\theta$. We write the $Q$-function under policy $\pi(s,a;\theta)$ as $Q^{\pi_{\theta}}(s,a)$, then we have the following theorem (for more details see \cite{suttonbarto}).
\begin{Theorem}[Policy Gradient Theorem \cite{sutton2000policy}]
Assume that $\pi(s,a;\theta)$ is differentiable with respect to $\theta$ and that there exists  $\mu^{\pi_{\theta}}$, the stationary distribution of the dynamics under policy $\pi_{\theta}$, which is independent of the initial state $s_0$. Then the policy gradient is
\begin{equation}\label{eqn:policy_gra_thm}
\nabla_{\theta}J(\theta) = \mathbb{E}_{s\sim \mu^{\pi_{\theta}},a\sim \pi_{\theta}}[\nabla_{\theta}\ln\pi(s,a;\theta)Q^{\pi_{\theta}}(s,a)].
\end{equation}
\end{Theorem}

For MDPs with finite state and action space, the stationary distribution exists under mild assumptions. For MDPs with infinite state and action space, more assumptions are needed, for example uniform geometric ergodicity \cite{konda2002}.

\subsubsection{REINFORCE: Monte Carlo Policy Gradient}\label{sec:REINFORCE}

Based on the policy gradient expression in \eqref{eqn:policy_gra_thm}, it is natural to estimate the expectation by averaging over samples of actual rewards, which is essentially the spirit of a standard \emph{Monte Carlo} method that repeatedly simulates a trajectory of length $M$ within each iteration. Now we introduce our first policy-based algorithm, called REINFORCE \cite{williams1992} or Monte Carlo policy gradient (see Algorithm \ref{alg:REINFORCE}). 
We use a simple empirical return function $G_t$ to approximate the $Q$-function in \eqref{eqn:policy_gra_thm}. The return $G_t$ is defined as the sum of the discounted rewards and given by $G_t:=\sum_{s=t+1}^{M} \gamma^{s-t-1}r_{s}$. Note that REINFORCE is a Monte Carlo algorithm since it employs the complete sample trajectories, that is, when estimating the return from time $t$, it uses all future rewards up until the end of the trajectory (see line 7 in Algorithm \ref{alg:REINFORCE}). Within the $n$-iteration the policy parameter $\theta^{(n+1)}$ is updated $M$ times with $G_t$ for $t=0,1,\cdots,M-1$. 

As a standard Monte Carlo method, REINFORCE provides an unbiased estimate of the policy gradient, however it suffers from high variance which therefore may lead to slow convergence. A popular variant of REINFORCE is to subtract a \emph{baseline}, which is a function of the state $s$, from the return $G_t$. This reduces the variance of the policy gradient estimate while keeping the mean of the estimate unchanged. A commonly used baseline is an estimate of the value function, $\widehat{V}(s)$, which can be learnt using one of the methods introduced in Section \ref{sec:value_based_methods}. Replacing the $G_t$ in \eqref{eqn:alg_REINFORCE_update} by $G_t-\widehat{V}(s)$ gives a REINFORCE algorithm with baseline.

\begin{algorithm}[H]
\caption{\textbf{REINFORCE: Monte Carlo Policy Gradient}}
\label{alg:REINFORCE}
\begin{algorithmic}[1]
    \STATE \textbf{Input}: total number of iterations $N$, learning rate $\beta$, a differentiable policy parametrization $\pi(s,a;\theta)$, finite length of the trajectory $M$.
    \STATE Initialize policy parameter $\theta^{(0)}$.
    \FOR {$n=0,1,\ldots,N-1$}
    \STATE Sample a trajectory $s_0,a_0,r_1,\cdots,s_{T-1},a_{T-1},r_T\sim\pi(s,a;\theta^{(n)})$
    \STATE Initialize $\theta^{(n+1)} = \theta^{(n)}$
    \FOR {$t=0,\cdots,M-1$}
    \STATE Calculate the return: $G_t=\sum_{s=t+1}^{M} \gamma^{s-t-1}r_{s}$
    \STATE Update the policy parameter
    \begin{equation}\label{eqn:alg_REINFORCE_update}
        \theta^{(n+1)} \leftarrow \theta^{(n+1)} + \beta\gamma^t G_t\nabla_{\theta}\ln\pi(s_t,a_t;\theta^{(n+1)})
    \end{equation}
    \ENDFOR
    \ENDFOR
\end{algorithmic}
\end{algorithm}

\subsubsection{Actor-Critic Methods}


The fact that REINFORCE provides unbiased policy estimates but may suffer from high variance is an example of the bias-variance dilemma (see, e.g.,  \cite{franccois2019} and \cite{von2011statistical}) which occurs in many machine learning problems. Also, as a Monte Carlo algorithm, REINFORCE requires complete trajectories so we need to wait until the end of each episode to collect the data. Actor-Critic methods can resolve the above issues by learning both the value function and the policy. The learned value function can improve the policy update through, for example, introducing bias into the estimation. Actor-Critic methods can also learn from incomplete experience in an online manner.

In Actor-Critic methods the value function is parametrized by $w$ and the policy is parametrized by $\theta$. These methods consist of two models: a \emph{critic} which updates the value function or $Q$-function parameter $w$, and an \emph{actor} which updates the policy parameter $\theta$ based on the information given by the critic. A natural idea is to use policy-based methods for the actor and use one of the methods introduced in Section \ref{sec:value_based_methods} for the critic; for example SARSA or $Q$-learning. We give the pseudocode for a simple Actor-Critic method in Algorithm \ref{alg:actor_critic}. 
Here for the critic, we approximate the $Q$-function by a linear combination of features, that is, $Q(s,a;w)=\phi(s,a)^\top w$ for some known feature $\phi:\mathcal{S}\times \mathcal{A}\rightarrow\mathbb{R}^d$, and use TD(0) to update the parameter $w$ to minimize the least-square error of the TD error with the gradient of the least-square error taking the form $\delta \phi(s,a)$ (see lines 9-10 in Algorithm \ref{alg:actor_critic}). For the actor, we use the (vanilla) policy-based method to update the policy parameter $\theta$ (see line 11 in Algorithm \ref{alg:actor_critic}).

There are three main ways to execute the algorithm. In the \emph{nested-loop} setting (see, e.g. \cite{kumar2019sample,xu2020improving}), the actor updates the policy in the outer loop after the critic's repeated updates in the inner loop. The second way is the \emph{two time-scale} setting (see, e.g. \cite{xu2020non}), where the actor and the critic update their parameters simultaneously with different learning rates. Note that the learning rate for the actor is typically much smaller than that of the critic in this setting. In the third way, the \emph{single-scale} setting, the actor and the critic update their parameters simultaneously but with a much larger learning rate for the actor than for the critic (see, e.g. \cite{fu2020single,xu2021doubly}).

\begin{algorithm}[H]
\caption{\textbf{Actor-Critic Algorithm}}
\label{alg:actor_critic}
\begin{algorithmic}[1]
    \STATE \textbf{Input}: A differentiable policy parametrization $\pi(s,a;\theta)$, a differentiable $Q$-function parameterization $Q(s,a;w)$, learning rates $\beta^{\theta}>0$ and $\beta^{w}>0$, number of iterations $N$
    \STATE Initialize policy parameter $\theta^{(0)}$ and $Q$-function parameter $w^{(0)}$
    \STATE Initialize $s$ 
    \FOR{$n=0,1,\ldots,N-1$}
    \STATE Sample $a\sim\pi(s,a;\theta^{(n)})$
    \STATE Take action $a$, observe state $s^\prime$ and reward $r$
    \STATE Sample action $a^\prime\sim\pi(s^\prime,\cdot;\theta^{(n)})$
    \STATE {Compute the TD error:} $\delta\leftarrow r+\gamma Q(s^\prime,a^\prime;w^{(n)})-Q(s,a;w^{(n)})$
    \STATE {Update $w$:} $w^{(n+1)}\leftarrow w^{(n)}+\beta^{w}\delta\phi(s,a)$ 
    \STATE $\theta^{(n+1)}\leftarrow \theta^{(n)} + \beta^{\theta}Q(s,a;w^{(n)})\nabla_{\theta}\ln\pi(s,a;\theta^{(n)})$
    \STATE $s\leftarrow s^\prime$, $a\leftarrow a^\prime$
    \ENDFOR
\end{algorithmic}
\end{algorithm}

\subsubsection{Discussion}\label{sec:value_discussion}
 
\paragraph{Variants of Policy-based Methods.} There have been many variations of policy-based methods with improvements in various directions. For example, the \emph{natural} policy-based algorithms \cite{kakade2001natural,peters2008natural} modify the search direction of the vanilla version by involving a \emph{Fisher information matrix} in the gradient ascent updating rule, which speeds the convergence significantly. To enhance the stability of learning, we may request the updated policy estimate to be not too far away from the previous estimate. Along this line,
TRPO \cite{schulman2015} uses the \emph{Kullback-Leibler (KL)-divergence} to measure the change between the consecutive updates as a constraint, while PPO \cite{schulman2017proximal} incorporate an \emph{adaptive KL-penalty} or a \emph{clipped objective} in the objective function to eliminate the constraint. When using the KL-penalty, PPO can be viewed as a Lagrangian relaxation of the TRPO algorithm. To reduce the sample complexity, Deterministic Policy Gradient (DPG) \cite{silver2014deterministic} uses a deterministic function (rather than the stochastic policy $\pi$) to represent the policy to avoid sampling over actions (although enough exploration needs to be guaranteed in other ways); Actor-Critic with Experience Replay (ACER) \cite{wang2016} reuses some experience (tuples of data collected in previous iterations/episodes, see the discussion of Deep $Q$-Networks in Section \ref{subsec:deep_value_based}) which improves the sample efficiency and decreases the data correlation. In addition we mention Asynchronous Advantage Actor-Critic (A3C) and Advantage Actor-Critic (A2C), two popular Actor-Critic methods with a special focus on parallel training \cite{mnih2016asynchronous}. The latter is a synchronous, deterministic version of the former. For continuous-time framework, see developments in \cite{jia2021policy,jia2022policy}.

\paragraph{Convergence Guarantees.} 
Now we discuss the convergence guarantees of policy-based algorithms. \cite{sutton2000policy} provided the first asymptotic convergence result for policy gradient methods with arbitrary differentiable function approximation for MDPs with bounded reward. They showed that such algorithms, including REINFORCE and Actor-Critic methods, converge to a locally stationary point.
For more examples of asymptotic convergence, see for example \cite{konda2000} and \cite{bhatnagar2010actor}.

Based on recent progress in non-convex optimization, non-asymptotic analysis of policy-based methods were first established for convergence to a stationary point. For example, \cite{kumar2019sample} provided a convergence rate analysis for a nested-loop Actor-Critic algorithm to the stationary point through quantifying the smallest number of actor updates $k$ required to attain $\inf_{0\leq m\leq k}\|\nabla J(\theta^{(k)})\|^2< \varepsilon$. We denote this smallest number as $K$. When the actor uses a policy gradient method, the critic achieves $K\leq \widetilde{\mathcal{O}}^{\prime}(1/\varepsilon^4)$ by employing TD(0), $K\leq \widetilde{\mathcal{O}}^{\prime}(1/\varepsilon^{3})$ by employing the Gradient Temporal Difference, and $K\leq \widetilde{\mathcal{O}}^{\prime}(1/\varepsilon^{5/2})$ by employing the Accelerated Gradient Temporal Difference, with continuous state and action spaces. \cite{zhang2020global} investigated a policy gradient method with Monte Carlo estimation of the policy gradient, called the RPG algorithm, where the rollout length of trajectories is drawn from a Geometric distribution. This generates unbiased estimates of the policy gradient with bounded variance, which was shown to converge to the first order stationary point with diminishing or constant learning rate at a sublinear convergence rate. They also proposed a periodically enlarged learning rate scheme and proved that the modified RPG algorithm with this scheme converges to the second order stationary point in a polynomial number of steps. 
For more examples of sample complexity analysis for convergence to a stationary point, see for example \cite{papini2018stochastic,xu2020improved,xu2019sample,shen2019hessian,xiong2020non}. The global optimality of stationary points was studied in \cite{bhandari2019} where they identified certain situations under which the policy gradient objective function has no sub-optimal stationary points despite being non-convex. 

Recently there have been results on the sample complexity analysis of the convergence to the global optimum.
\cite{cen2020fast} proved that the entropy regularized natural policy gradient method achieves a $Q$-sample complexity (and $\varepsilon$-optimal policies) with linear convergence rate. \cite{shani2020adaptive} showed that with high probability at least $1-\delta$, the TRPO algorithm has sublinear rate $ \widetilde{\mathcal{O}}^{\prime}(1/\varepsilon^2)$ with mean-squared sample complexity $\widetilde{\mathcal{O}}\big(\frac{|\mathcal{A}|^2(|\mathcal{S}|+\log(1/\delta))}{(1-\gamma)^3\varepsilon^4}\big)$ in the unregularized (tabular) MDPs, and has an improved rate $\widetilde{\mathcal{O}}^{\prime}(1/\varepsilon)$ with mean-squared sample complexity $\widetilde{\mathcal{O}}\big(\frac{|\mathcal{A}|^2(|\mathcal{S}|+\log(1/\delta))}{(1-\gamma)^4\varepsilon^3}\big)$ in MDPs with entropy regularization.  
\cite{xu2020non} gave the first non-asymptotic convergence guarantee for the two time-scale natural Actor-Critic algorithms with mean-squared sample complexity of order $\widetilde{\mathcal{O}}(\frac{1}{(1-\gamma)^9\varepsilon^{4}})$. For single-scale Actor-Critic methods, the global convergence with sublinear convergence rate was established in both \cite{fu2020single} and \cite{xu2021doubly}. The non-asymptotic convergence of policy-based algorithms is shown in other settings, see \cite{zhang2020sample} for the regret analysis of the REINFORCE algorithm for discounted MDPs and \cite{agarwal2021theory,mei2020} for policy gradient methods in the setting of known model parameters. 


\subsection{General Discussion}\label{sec:discussion}
So far we have focused on model-free RL with discounting over an infinite time horizon. In this section, we discuss two other cases: RL over a finite time horizon with both model-based and model-free methods, and model-based RL with an infinite time horizon.

\subsubsection{RL with Finite Time Horizon}
As discussed earlier, episodic RL with finite time horizon has also been widely used in many financial applications. In this section, we discuss some  episodic RL algorithms (with both model-based and model-free methods) and their performance through both sample complexity and regret analysis.
\cite{PSRL2013} proposed a model-based algorithm, known as Posterior Sampling for Reinforcement Learning (PSRL) which is a model-based algorithm, and establishes an $\widetilde{\mathcal{O}}(T|\mathcal{S}|\sqrt{|\mathcal{A}|M})$ expected regret bound.  \cite{DannBrunskill2015} proposed a so-called Upper-Confidence Fixed-Horizon (UCFH) RL algorithm that is model-based and has $V$-sample complexity\footnote[2]{In the original papers the sample complexity is defined in terms of the number of episodes. Here we multiply the original order in these papers by $T$ to match the definition of sample complexity in this paper.} of order $\widetilde{\mathcal{O}}(\frac{|\mathcal{S}|^2|\mathcal{A}|\,T^3}{\varepsilon^2})$ assuming known rewards.
\cite{azar2017minimax} considered two model-based algorithms in the setting of known reward, called the UCBVI-CH and UCBVI-BF algorithms, which were proved to achieve a regret bound of $\widetilde{\mathcal{O}}(T\sqrt{|\mathcal{S}|\,|\mathcal{A}|M})$ (when $M\geq T|\mathcal{S}|^3|\mathcal{A}|$ and $|\mathcal{S}||\mathcal{A}|\geq T$) and $\widetilde{\mathcal{O}}(\sqrt{T|\mathcal{S}|\,|\mathcal{A}|M})$ (when $M\geq T^3|\mathcal{S}|^3|\mathcal{A}|$ and $|\mathcal{S}||\mathcal{A}|\geq T$), respectively.
\cite{DannUBEV} proposed an Upper Bounding the Expected Next State Value (UBEV) algorithm which achieves a sample complexity\footnotemark[2] of $\widetilde{\mathcal{O}}\big(\frac{|\mathcal{S}|^2|\mathcal{A}|\,T^5}{\varepsilon^2}{\rm polylog}(\frac{1}{\delta})\big)$.
They also proved that the algorithm
has a regret bound of  $\widetilde{\mathcal{O}}(T^2\sqrt{|\mathcal{S}|\,|\mathcal{A}|M})$ with $M\ge |\mathcal{S}|^5|\mathcal{A}|^3$. \cite{jin2018q} proved that $Q$-learning with UCB exploration achieves regret of $\widetilde{\mathcal{O}}(\sqrt{T^3|\mathcal{S}|\,|\mathcal{A}|M})$ without requiring access to a simulator. The above regret bounds depend on the size of the state and action space and thus may suffer from the curse of dimensionality as $|\mathcal{S}|$ and $|\mathcal{A}|$ increase. To tackle this issue, \cite{yang2020reinforcement} learned a low-dimensional representation of the probability transition model and proved that their MatrixRL algorithm achieves a regret bound of $\widetilde{\mathcal{O}}(T^2\,d\sqrt{M})$ which depends on the number of features $d$ rather than $|\mathcal{S}|$ and $|\mathcal{A}|$.
\cite{jin2020provably} provided a  $\widetilde{\mathcal{O}}(\sqrt{d^3T^3M})$ regret bound with high probability for a modified version of the Least-Squares Value Iteration (LSVI) algorithm with UCB exploration.

\subsubsection{Model-based RL with Infinite Time Horizon}

To derive policies by utilizing  estimated transition probabilities and reward functions under the infinite time horizon setting, \cite{brafman2002r} proposed an R-MAX algorithm which can also be applied in zero-sum stochastic games. The R-MAX algorithm was proved to achieve a sample complexity of exploration of order $\widetilde{\mathcal{O}}\big(\frac{|\mathcal{S}|^2|\mathcal{A}|}{(1-\lambda)^6\varepsilon^3}\big)$ by \cite{li2009rmax}.  The Model-based Interval Estimation (MBIE) in \cite{Strehl2005} was proved to achieve a sample complexity of exploration of the same order as R-MAX. \cite{MoRmax2010} further modified the R-MAX algorithm to an MoRmax algorithm with an improved sample complexity of exploration of order $\widetilde{\mathcal{O}}\big(\frac{|\mathcal{S}|\,|\mathcal{A}|}{(1-\lambda)^6\varepsilon^2}\big)$.
\cite{azar2013minimax} proved that the model-based $Q$-Value Iteration (QVI) achieves the $Q$-sample complexity of $\widetilde{\mathcal{O}}(\frac{|\mathcal{S}|\,|\mathcal{A}|}{(1-\gamma)^3\varepsilon^2)})$ for some small $\varepsilon$
with high probability at least $1-\delta$. \cite{agarwal2020model} studied the approximate MDP model with any accurate black-box planning algorithm and showed that this has the same sample complexity as \cite{azar2013minimax}, but with a larger range of $\varepsilon$.
See also \cite{strehl2009reinforcement,kearns2002near,lattimore2012pac} for model-based RL along this line. 

\emph{Linear-quadratic} (LQ) problems are another type of model-based RL problem that assumes linear state dynamics and quadratic objective functions with continuous state and action space. These problems are one of the few optimal control problems for which there exists a closed-form analytical representation of the optimal feedback control. Thus LQ frameworks, including the (single-agent) \emph{linear-quadratic regulator} (LQR) and (multi-agent) LQ games, can be used in a wide variety of applications, such as the optimal execution problems which we will discuss later. For a theoretical analysis of RL algorithms within the LQ framework, see for example \cite{fazel2018global} and \cite{hambly2020policy} for the global linear convergence of policy gradient methods in LQR problems with infinite and finite horizon, respectively. See also \cite{basei2021logarithmic} and \cite{guo2021reinforcement}  for the regret analysis of  linear-quadratic reinforcement learning algorithms  in continuous time.

\subsubsection{Exploration Exploitation Trade-offs}\label{sec:ex-ex}
As mentioned earlier, exploitation versus exploration is a critical topic in RL. RL agents want to find the optimal solution as fast as possible, meanwhile committing to solutions too fast without enough exploration could lead to local minima or total failure. Modern RL algorithms that optimize for the best returns can achieve good exploitation quite efficiently, while efficient exploration remains a challenging and less understood topic.

For multi-armed bandit problems or MDP with finite state and action spaces, classic exploration algorithms such as $\varepsilon$-greedy, UCB, Boltzmann exploration and Thompson sampling show promising performance in different settings.
For the $\varepsilon$-greedy algorithm \cite{dann2022guarantees,dabney2020temporally}, the agent  randomly explores occasionally with probability $\varepsilon$ and takes the optimal action most of the time with probability $1-\varepsilon$. For example,  see Equation \eqref{epsilon_policy} for the $\varepsilon$-greedy method combined with a value-based algorithm. The UCB type of algorithms are mainly designed for value-based algorithms \cite{dong2019q,azar2017minimax,jin2018q,jin2020provably}.  The agent selects the greediest action to maximize the upper confidence bound  $Q(s,a) + U(s,a)$, where $Q(s,a)$
 is the Q value function  and $U(s,a)$
 is a function inversely proportional to how many times the state-action pair $(s,a)$  has been taken (hence proportional to the agent's confidence in her estimation of the Q function). For Boltzmann exploration \cite{asadi2017alternative,haarnoja2017reinforcement,WZZ2018}, the agent draws actions from a Boltzmann distribution (softmax) over the learned Q value function, regulated by a temperature parameter. For the Thompson sampling method \cite{thompson1933likelihood,ouyang2017learning,gopalan2015thompson}, the agent keeps track of a  {\it belief of the optimal action} via a distribution over the set of admissible actions (for each given state).

For deep RL training when neural networks are used for function approximation, entropy regularization (adding an entropy term  into the loss function) and noise-based exploration (adding noise into the observation, action or even parameter space) are often used for better exploration of the parameter space.


\subsubsection{Regularization in Reinforcement Learning}\label{subsec:regularization}
Many recent developments in RL benefit from regularization, which has been shown to improve not only the exploration but also the robustness. Examples include both policy-based methods and value-based methods in various settings. Here we  provide a general overview of different regularization techniques and their advantages along with several examples. For example,
TRPO and PPO use the KL-divergence between two consecutive policies as a penalty in policy improvement \cite{schulman2015,schulman2017proximal}. Soft-$Q$-learning uses {\it Shannon entropy} as a penalty in value iteration \cite{haarnoja2017reinforcement}. The authors confirmed in simulated experiments that entropy regularization helps to improve exploration and allows transferring skills between tasks. \cite{geist2019theory} proposed a unified framework to analyze the above algorithms via a regularized Bellman operator. 
\cite{cen2020fast} showed that entropy regularization can also help to speed up the convergence rate from a theoretical perspective.
It is worth noting that \cite{WZZ2018} explained the exploitation-exploration trade-off with entropy regularization from a continuous-time stochastic control perspective and provided theoretical support for Gaussian exploration from the special case of linear-quadratic regulator. Recently \cite{gao2020state} and \cite{tang2021exploratory} extended this idea to a general control framework beyond the linear-quadratic case. \cite{grau2018soft} extended the idea of Shannon's entropy regularization to mutual-information regularization and showed its superior performance when actions have significantly different importance.
When functional approximations are applied in RL training, the regularization technique is shown to be a powerful, flexible, and  principled way to balance approximation and estimation errors \cite{massoud2009regularized,farahmand2008regularized}. The idea of regularization is that one starts from a large function space and controls the solution's complexity by a regularization (or penalization) term. Examples of regularization include the Hilbert norm (for $Q$-functions) and the $l_1$ norm (for parameters).


\subsubsection{Reinforcement Learning in Non-stationary Environment}
Most existing work on reinforcement learning considers a stationary environment and aims to find the optimal policy or a policy with low (static) regret. In many financial applications, however, the environment is far from being stationary. We introduce
the mathematical formulations of non-stationary RL (in both episodic and infinite horizon settings) below. 

\paragraph{Non-stationary Episodic RL.} For the episodic (or finite-time horizon) setting, an agent interacts with a non-stationary MDP
for $K$ episodes, with each episode containing $T$ timestamps. Let $(t, k)$ denote a (time,episode)
index for the $t$-th timestamp in the $k$-th episode. The non-stationary environment can be denoted by a tuple
$(\mathcal{S},\mathcal{A}, T, \pmb{P}, \pmb{r})$, where $\mathcal{S}$ is the state space, $\mathcal{A}$ is the action space, $T$ is the number of timestamps in one episode, $\pmb{P} = \{P_{k,t} \}_{1 \leq k\leq K, 0\leq t\leq T}$
is the set of transition kernels with $P_{k,t}: \mathcal{S}\times \mathcal{A}\rightarrow\mathcal{P}(\mathcal{S})$ for all $1\leq k\leq K$ and $0\leq t\leq T$,
and $\pmb{r} = \{r_{k,t} \}_{1 \leq k\leq K, 0\leq t\leq T}$ is the set of reward functions with $r_{k,t}(s,a)$ a random variable in $\mathbb{R}$ (depending on both the time-step and the episode number) for each pair $(s,a)\in \mathcal{S}\times \mathcal{A}$ and for $0\leq t\leq T-1$. Similarly, the terminal reward $r_{k,T}(s)$ is a real-valued random variable for each $s\in\mathcal{S}$. Denote by $\Pi=\{\pi_{k,t} \}_{1 \leq k\leq K, 0\leq t\leq T-1}$ the Markovian policy in which $\pi_{k,t}$ can be either deterministic or randomized. The value function for this policy at the $k$-th episode can be written as
\begin{eqnarray}
V^{\Pi}_{k,t}(s) = \left.\mathbb{E}^{\Pi}\left[\sum_{u=t}^{T-1} r_{k,u}(s_u,a_u)+r_{k,T}(s_T)\right|s_t = s\right],
\end{eqnarray}
where $s_{u+1} \sim P_{k,u}(s_u,a_u)$ and $a_u\sim \pi_{k,u}(s_u)$.
\paragraph{Non-stationary RL with Infinite Time Horizon.}
For the infinite horizon setting, we drop the index indicating episodes and let the finite horizon $T$ become infinite. Then the non-stationary environment can be denoted by a tuple
$(\mathcal{S},\mathcal{A},  \pmb{P}, \pmb{r})$, with the obvious changes from the episodic case. Again we write $\Pi=\{\pi_{t} \}_{t \ge 0}$ for the Markovian policy in which $\pi_{t}$ can be either deterministic or randomized and consider the associated value function with a discount factor $\gamma$:
\begin{eqnarray}
V^{\Pi}_{t}(s) = \left.\mathbb{E}^{\Pi}\left[\sum_{u=t}^\infty \gamma^{u-t} r_{u}(s_u,a_u)\right|s_t = s\right],
\end{eqnarray}
where $s_{u+1} \sim P_{u}(\cdot|s_u,a_u)$ and $a_u\sim \pi_{u}(s_u)$.\\

Indeed, there has been a recent surge of theoretical studies on this topic for various settings such as infinite-horizon  MDPs with finite state and action spaces \cite{gajane2018sliding,cheung2019learning}, episodic MDPs with finite state and action spaces \cite{mao2020model}, 
and episodic linear MDPs \cite{touati2020efficient}.
 However, all these papers assume the agent has prior knowledge of the degree of nonstationarity such as the number of changes in the environment, which is not very practical for many applications.
 To relax this assumption, \cite{cheung2020reinforcement} proposes a Bandit-over-Reinforcement-Learning (BoRL) algorithm to extend their earlier result \cite{cheung2019learning}. However, it introduces extra overheads and leads to suboptimal regret.
 More recently, \cite{wei2021non} proposes a general approach that is applicable to various reinforcement learning settings
(including both episodic MDPs and infinite-horizon MDPs) and achieves optimal dynamic regret without any prior knowledge of the degree of non-stationarity.

\section{Deep Reinforcement Learning}\label{sec:deep_value_based}

In the setting where the state and action spaces are very large, or high dimensional or they are continuous it is often challenging to apply classical value-based and policy-based methods.
In this context, parametrized value functions ($Q(s,a;\theta)$ and $V(s,a;\theta)$)  or policies ($\pi(s,a;\theta)$) are more practical. Here we focus on neural-network-based parametrizations and functional approximations, which have the following advantages:
\begin{itemize}
    \item Neural networks are well suited for dealing with high-dimensional 
    inputs such as time series data, and, in practice, they do not require an exponential increase in data when adding extra dimensions to the state or action space.
\item  In addition, they can be trained incrementally and make use of additional samples obtained as learning happens.
\end{itemize}

In this section, we will introduce deep RL methods in the setting of an infinite time horizon with discounting, finite state and action spaces ($|\mathcal{A}|<\infty$ and $|\mathcal{S}|<\infty$), and stationary policies. 

\subsection{Neural Networks}

A typical neural network is a nonlinear function, represented by a collection of  neurons. These are typically arranged as a number of layers connected by operators such as filters, poolings and gates, that map an input variable in $\mathbb{R}^{n} $  to an output variable in $\mathbb{R}^{m}$ for some $n,m \in \mathbb{Z}^+$. 

In this subsection, we introduce three popular neural network architectures including fully-connected neural networks, convolutional neural networks \cite{lecun1995CNN} and recurrent neural networks \cite{sutskever2011RNN}.

\paragraph{Fully-connected Neural Networks.} A fully-connected Neural Network (FNN) is the simplest neural network architecture where any given neuron is connected to all neurons in the previous layer. To describe the setup we fix the number of layers $L \in \mathbb{Z}^{+}$ in the neural  network and the width of the $l$-th layer $n_l \in \mathbb{Z}^+$ for  $l=1,2,\ldots, L$. Then for an input variable  $\mathbf{z} \in \mathbb{R}^{n}$, the  functional form of the FNN is
\begin{eqnarray}\label{eq:generator}
F(\mathbf{z};\pmb{W},\pmb{b}) = \mathbf{W}_L \cdot \sigma \left( \mathbf{W}_{L-1} \cdots \sigma (\mathbf{W}_1 \mathbf{z}+\mathbf{b}_1)\cdots +\mathbf{b}_{L-1}\right) +\mathbf{b}_L,
\end{eqnarray}
in which $(\mathbf{W},\mathbf{b})$ represents all the parameters in the  neural network, with $\mathbf{W} = (\mathbf{W}_1,\mathbf{W}_2,\ldots,\mathbf{W}_L)$ and  $\mathbf{b} = (\mathbf{b}_1,\mathbf{b}_2,\ldots,\mathbf{b}_L)$. 
Here $\mathbf{W}_l \in \mathbb{R}^{n_l \times n_{l-1}}$ and  $\mathbf{b}_l \in \mathbb{R}^{n_l \times 1}$ for $l=1,2,\ldots,L$, where $n_0 = n$ is set as the dimension of the input variable.
The operator $\sigma(\cdot)$ takes a vector of any dimension as input, and applies a function component-wise.  Specifically, for any $q\in \mathbb{Z}^+$ and any vector $\mathbf{u}=(u_1,u_2,\cdots,u_q)^{\top} \in \mathbb{R}^{q}$, we have that 
\begin{eqnarray*}
\sigma(\mathbf{u}) = (\sigma(u_1),\sigma(u_2),\cdots,\sigma(u_q))^{\top}.
\end{eqnarray*} 
In the neural networks literature, the $\mathbf{W}_l$'s are often called the {\it weight} matrices, the $\mathbf{b}_l$'s are called {\it bias} vectors, and $\sigma(\cdot)$ is referred to as the {\it activation function}. Several popular choices for the activation function include ReLU with $\sigma(u) = \max(u,0)$, Leaky ReLU with $\sigma(u) = a_1\,\max(u,0) - a_2\, \max(-u,0)$ where $a_1,a_2>0$, and smooth functions such as $\sigma(\cdot) = \tanh(\cdot)$. In \eqref{eq:generator}, information propagates from the input layer to the output layer in a  feed-forward manner in the sense that connections between the nodes do not form a cycle. Hence \eqref{eq:generator} is also referred to as fully-connected feed-forward neural network in the deep learning literature.

\paragraph{Convolutional Neural Networks.} In addition to the FNNs above, convolutional neural networks (CNNs) are another type of feed-foward neural network that are especially popular for image processing. In the finance setting CNNs have been successfully applied  to price prediction based on inputs which are images containing visualizations of price dynamics and trading volumes \cite{jiang2020re}. The CNNs have two main building blocks -- convolutional layers and pooling layers. Convolutional layers are used to capture local patterns in the images and pooling layers are used to reduce the dimension of the problem and improve the computational efficiency.

Each convolutional layer uses a number of trainable \emph{filters} to extract features from the data. We start with the simple case of a single filter $\pmb{H}$, which applies the following convolution operation  $\pmb{z}\ast \pmb{H}:\mathbb{R}^{n_x\times n_y}\times \mathbb{R}^{k_x\times k_y}\rightarrow \mathbb{R}^{(n_x-k_x+1)\times(n_y-k_y+1)}$ to the input $\pmb{z}$ through
\[
[\pmb{z}\ast \pmb{H}]_{i,j} = \sum_{i^\prime=1}^{k_x}\sum_{j^\prime=1}^{k_y}[\pmb{z}]_{i+i^\prime-1,j+j^\prime-1}[\pmb{H}]_{i^\prime,j^\prime}.
\]
The output of the convolutional layer $\widehat{\pmb{z}}$ is followed by the activation function $\sigma(\cdot)$, that is
\[
[\widehat{\pmb{z}}]_{i,j}=\sigma([\pmb{z}\ast \pmb{H}]_{i,j}).
\]
The weights and bias introduced for FNNs can also be incorporated in this setting thus, in summary, the above simple CNN can be represented by
\[
F(\pmb{z};\pmb{H},\pmb{W},\pmb{b}) = \pmb{W}\cdot\sigma(\pmb{z}\ast \pmb{H})+\pmb{b}.
\]
This simple convolutional layer can be generalized to the case of multiple channels with multiple filters, where the input is a 3D tensor $\pmb{z}\in\mathbb{R}^{n_x\times n_y\times n_z}$ where $n_z$ denotes the number of channels. Each channel represents a feature of the input, for example, an image as an input typically has three channels, red, green, and blue. Then each filter is also represented by a 3D tensor $\pmb{H}_k\in\mathbb{R}^{k_x\times k_y\times n_z}$ ($k=1,\ldots,K$) with the same third dimension $n_z$ as the input and $K$ is the total number of filters. In this case, the convolution operation becomes
\[
[\pmb{z}\ast \pmb{H}_k]_{i,j} = \sum_{i^\prime=1}^{k_x}\sum_{j^\prime=1}^{k_y}\sum_{l=1}^{n_z}[\pmb{z}]_{i+i^\prime-1,j+j^\prime-1,l}[\pmb{H}]_{i^\prime,j^\prime,l},
\]
and the output is given by
\[
[\widehat{\pmb{z}}]_{i,j,k}=\sigma([\pmb{z}\ast \pmb{H}_k]_{i,j}).
\]

Pooling layers are used to aggregate the information and reduce the computational cost, typically after the convolutional layers. A commonly used pooling layer is the max pooling layer, which computes the maximum value of a small neighbourhood in the spatial coordinates. Note that in addition,  fully-connected layers, as introduced above, are often used in the last few layers of a CNN.

\paragraph{Recurrent Neural Networks.} Recurrent Neural Networks (RNNs) are a family of neural networks that are widely used in processing sequential data, including speech, text and financial time series data.
Unlike feed-forward neural networks, RNNs are a class of artificial neural networks where connections between units form a {\it directed cycle}. RNNs can use their internal memory to process arbitrary sequences of inputs and hence  are applicable to tasks such as sequential data processing.

For RNNs, our input is a sequence of data $\pmb{z}_1,\pmb{z}_2,\ldots,\pmb{z}_T$. An RNN models the \emph{internal} state $\pmb{h}_t$ by a recursive relation 
\[
\pmb{h}_t = F(\pmb{h}_{t-1},\pmb{z}_t;\theta),
\]
where $F$ is a neural network with parameter $\theta$ (for example $\theta=(\pmb{W},\pmb{b})$). Then the output is given by
\[
\widehat{\pmb{z}}_t = G(\pmb{h}_{t-1},\pmb{z}_t;\phi),
\]
where $G$ is another neural network with parameter $\phi$ ($\phi$ can be the same as $\theta$).
There are two important variants of the vanilla RNN introduced above -- the long short-term memory (LSTM) and the gated recurrent units (GRUs).  Compared to vanilla RNNs, LSTM and GRUs are shown to have better performance in handling sequential data with long-term dependence due to their flexibility in propagating information flows. Here we introduce the LSTM. Let $\odot$ denote element-wise multiplication. The LSTM network architecture is given by
\begin{eqnarray*}
\pmb{f}_t &=& \sigma(\pmb{U}^{f}\pmb{z}_t+\pmb{W}^f \pmb{h}_{t-1}+\pmb{b}^f) \\
\pmb{g}_t &=& \sigma(\pmb{U}^{g}\pmb{z}_t+\pmb{W}^g \pmb{h}_{t-1}+\pmb{b}^g) \\
\pmb{q}_t &=& \sigma(\pmb{U}^{q}\pmb{z}_t+\pmb{W}^q \pmb{h}_{t-1}+\pmb{b}^q) \\
\pmb{c}_t &=& \pmb{f}_t\odot \pmb{c}_{t-1}+\pmb{g}_t\odot\sigma(\pmb{U}\pmb{z}_t+\pmb{W}\pmb{h}_{t-1}+\pmb{b}) \\
\pmb{h}_t &=& \tanh(\pmb{c}_t)\odot \pmb{q}_t 
\end{eqnarray*}
where $\pmb{U}, \pmb{U}^f, \pmb{U}^g, \pmb{U}^q, \pmb{W}, \pmb{W}^f, \pmb{W}^g, \pmb{W}^q$ are trainable weights matrices, and $\pmb{b}, \pmb{b}^f, \pmb{b}^g, \pmb{b}^q$ are trainable bias vectors. In addition to the internal state $\pmb{h}_t$, the LSTM also uses the \emph{gates} and a \emph{cell} state $\pmb{c}_t$. The forget gate $\pmb{f}_t$ and the input gate $\pmb{g}_t$ determine how much information can be transmitted from the previous cell state $\pmb{c}_{t-1}$ and from the update $\sigma(\pmb{U}\pmb{z}_t+\pmb{W}\pmb{h}_{t-1}+\pmb{b})$, to the current cell state $\pmb{c}_t$. The output gate $\pmb{q}_t$ controls how much $\pmb{c}_t$ reveals to $\pmb{h}_t$. For more details see \cite{sutskever2011RNN} and \cite{fan2019selective}. 

\paragraph{Training of Neural Networks.} Mini-batch stochastic gradient descent is a popular choice  for training neural networks due to its sample and computational efficiency. In this approach the parameters $\theta=(\pmb{W},\pmb{b})$ are updated in the descent direction of an objective function $\mathcal{L}(\theta)$ by selecting a mini-batch of samples at random to estimate the gradient of the objective function $\nabla_{\theta}\mathcal{L}(\theta)$ with respect to the parameter $\theta$. For example, in supervised learning, we aim to learn the relationship between an input $X$ and an output $Y$, and the objective function $\mathcal{L}(\theta)$ measures the distance between the model prediction (output) and the actual observation $Y$. Assume that the dataset contains $M$ samples of $(X,Y)$. Then the mini-batch stochastic gradient descent method is given as
\begin{equation}\label{eqn:mini_batch_gd}
    \theta^{(n+1)} = \theta^{(n)} - \beta \widehat{\nabla_{\theta}\mathcal{L}(\theta^{(n)})}, \qquad n=0,\ldots,N-1,
\end{equation}
where $\beta$ is a constant learning rate and $\widehat{\nabla_{\theta}\mathcal{L}(\theta^{(n)})}$ is an estimate of the true gradient $\nabla_{\theta}\mathcal{L}(\theta^{(n)})$, which is computed by averaging over $m$ ($m\ll M$) samples of $(X,Y)$.  It is called (vanilla) stochastic gradient descent when $m=1$,and it is the (traditional) gradient descent when $m=M$. Compared with gradient descent, mini-batch stochastic gradient descent is noisy but computationally efficient since it only uses a subset of the data, which is advantageous when dealing with large datasets. It is also worth noting that, in addition to the standard mini-batch stochastic gradient descent methods \eqref{eqn:mini_batch_gd}, momentum methods are popular extensions which take into account the past gradient updates in order to accelerate the learning process. We give the updating rules for two examples of gradient descent methods with momentum -- the standard momentum and the Nesterov momentum \cite{nesterov1983},
\begin{equation*}
 \begin{cases}
     \underbrace{z_{n+1}=\beta z_{n}+\nabla_{\theta}\mathcal{L}(\theta^{(n)})}_{\text{the standard momentum}}\quad \text{or} \quad \underbrace{z_{n+1}=\beta z_{n}+\nabla_{\theta} \mathcal{L}(\theta^{(n)}-\alpha z_{n})}_{\text{the Nesterov momentum}}\\
     \theta^{(n+1)}=\theta^{(n)}-\alpha z_{n+1},
    \end{cases}
\end{equation*}
where $\alpha$ and $\beta$ are constant learning rates. Such methods are particularly useful when the algorithms enter into a region where the gradient changes dramatically and thus the learned parameters can bounce around the region which slows down the progress of the search.
Additionally, there are many other variants such as RMSprop \cite{tieleman2012lecture} and ADAM \cite{kingma2014adam}, both of which employ adaptive learning rates.

\subsection{Deep Value-based RL Algorithms}\label{subsec:deep_value_based}

In this section, we introduce several $Q$-learning algorithms with neural network approximations. We refer  the reader to \cite{franccois2018introduction} for other deep value-based RL algorithms such as Neural TD learning and Dueling $Q$-Networks.

\paragraph{Neural Fitted $Q$-learning.}

Fitted $Q$-learning \cite{gordon1996stable} is a generalization of the classical $Q$-learning algorithm with functional approximations and it is applied in an off-line setting with a pre-collected dataset in the form of tuples $(s, a, r, s')$ with $s'\sim P(s,a)$ and $r = r(s,a)$ which is random. When the class of approximation functionals is constrained to neural networks, the algorithm is referred to as Neural Fitted $Q$-learning  and the $Q$-function is parameterized by   $Q(s,a;\theta) = F((s,a);\theta)$ with $F$ a neural network function parameterized by $\theta$ \cite{riedmiller2005neural}. For example, $F$ can be set as \eqref{eq:generator} with $\theta = (\pmb{W},\pmb{b})$.

In Neural Fitted $Q$-learning, the algorithm starts with some random initialization of the $Q$-values $Q(s, a; \theta^{(0)})$ where $\theta^{(0)}$ refers to the initial parameters. Then, an approximation of the $Q$-values at the $n$-th iteration $Q(s,a;\theta^{(n)})$ is updated towards the target value
\begin{eqnarray}\label{eq:Y_update}
Y^Q_n = r + \gamma \max_{a'\in \mathcal{A}}Q(s',a';\theta^{(n)}),
\end{eqnarray}
where $\theta^{(n)}$ are the neural network parameters in the $n$-th iteration, updated by stochastic gradient descent (or a variant) by minimizing the square loss:
\begin{eqnarray}
\mathcal{L}_{NFQ}(\theta^{(n)}) = \|Q(s,a;\theta^{(n)})-Y^Q_n\|_2^2.
\end{eqnarray}
Thus, the $Q$-learning update amounts to updating the parameters:
\begin{eqnarray}\label{eq:deep_q_update}
\theta^{(n+1)} = \theta^{(n)} +\beta \left(Y_n^Q-Q(s,a;\theta^{(n)})\right) \nabla_{\theta^{(n)}} Q(s,a;\theta^{(n)})
\end{eqnarray}
where $\beta$ is a learning rate. This update resembles stochastic gradient descent, updating the current value $Q(s, a; \theta^{(n)})$ towards the target value $Y_n^Q$.

When neural networks are applied to approximate the $Q$-function, it has been empirically observed that Neural Fitted $Q$-learning may suffer from slow convergence or even divergence \cite{riedmiller2005neural}. In addition, the approximated $Q$-values tend to be overestimated due to the max operator \cite{van2016deep}. 

\paragraph{Deep $Q$-Network (DQN).}
To overcome the instability issue
and the risk of overestimation mentioned above, \cite{mnih2015human} proposed a Deep $Q$-Network (DQN) algorithm in an online setting with two novel ideas.  One idea is the slow-update of the target network
and the  second is the use of `experience replay'.  Both these ideas dramatically improve the empirical  performance of the algorithm and DQN has been shown to have a strong performance for a variety of ATARI  games \cite{bellemare2013arcade}. 

We first discuss the use of experience replay \cite{lin1992self}. That is we introduce a replay memory $\mathcal{B}$. At each time $t$, the tuple  $(s_t, a_t, r_t,s_{t+1})$ is stored in $\mathcal{B}$ and a mini-batch of {$B$} independent samples  is randomly selected from $\mathcal{B}$ to train the neural network via stochastic gradient descent. Since
the trajectory of an MDP has strong temporal correlation, the goal of experience replay is to obtain
uncorrelated samples that are more similar to the i.i.d data (often assumed in many optimization algorithms), which can give more accurate gradient estimation  for the stochastic optimization problem and enjoy better convergence performance. For experience replay,  the replay memory size is usually very large in practice. For example, the replay memory size is $10^6$
in \cite{mnih2015human}. Moreover, DQN uses the $\varepsilon$-greedy policy, which enables exploration over the state-action space
$\mathcal{S}\times \mathcal{A}$. Thus, when the replay memory is large, experience replay is close to sampling independent transitions from an explorative policy. This reduces the variance of the 
gradient 
which is used to update $\theta$. Thus, experience replay stabilizes the training of DQN, which benefits the algorithm in terms of computational efficiency.


We now discuss using a target network $Q(\cdot,\cdot;{\theta}^{-})$ with parameter ${\theta}^{-}$ (the current estimate of the parameter). With independent samples $\{(s_{(i)}, a_{(i)}, r_{(i)}, s_{(i)}^{\prime})\}_{i=0}^{B}$ 
from the replay memory (we use $s_{(i)}^{\prime}$
instead of $s_{(i+1)}$ for the state after $(s_{(i)},a_{(i)})$ to avoid notational confusion with the next independent sample $s_{(i+1)}$ in the state space), to update the parameter $\theta$ of the $Q$-network, we compute the target 
\begin{eqnarray}\label{eq:Y_update2}
\widetilde{Y}_i^Q = r_{(i)} +  \gamma \max_{a'\in\mathcal{A}}Q(s_{(i)}',a';\theta^{(n)-}),
\end{eqnarray}
 and update $\theta$ using the gradient of
\begin{eqnarray} 
\mathcal{L}_{DQN}(\theta^{(n)},\theta^{(n)-}) =\frac{1}{B}\sum_{i=1}^B \left\|Q(s_{(i)},a_{(i)};\theta^{(n)})-\widetilde{Y}_i^Q\right\|_2^2.
\end{eqnarray}
Whereas the parameter $\theta^{(n)-}$ is updated once every $T_{\rm target}$ steps by letting $\theta^{(n)-} = \theta^{(n)}$, if $n = m T_{\rm target}$ for some $m\in \mathbb{Z}^{+}$, and $\theta^{(n)-} = \theta^{(n-1)-}$ otherwise. That is, the target network is held fixed for $T_{\rm target}$ steps and then updated using the current weights of the $Q$-network. The introduction of a target network prevents the rapid propagation of instabilities and it reduces the risk of divergence 
The idea of target networks can be seen as an instantiation of Fitted $Q$-learning, where each period between target network updates corresponds to a single Fitted $Q$-iteration.
\paragraph{Double Deep $Q$-network (DQN).} The max operator in Neural Fitted $Q$-learning and DQN, in \eqref{eq:Y_update}
and \eqref{eq:Y_update2}, uses the same values both to select and to evaluate
an action. This makes it more likely to select overestimated
values, resulting in over optimistic value estimates. To prevent this, Double $Q$-learning \cite{hasselt2010double} decouples the selection from the evaluation and this is further extended to the neural network setting \cite{van2016deep}.

In double $Q$-learning \cite{hasselt2010double} and double deep $Q$-network \cite{van2016deep}, two value functions are learned by assigning experiences randomly to update one of the two value functions, resulting in two sets of weights, $\theta$ and $\eta$. For each update, one set of weights is used to determine the greedy policy and the other to determine its value. For a clear comparison, we can untangle the
selection and evaluation in Neural Fitted $Q$-learning and rewrite its target  \eqref{eq:Y_update} as
\begin{eqnarray}\label{eq:Y_update_prime}
Y^Q_n = r + \gamma\, Q(s',\arg\max_{a\in \mathcal{A}}Q(s',a;\theta^{(n)});\theta^{(n)}).
\end{eqnarray}
The target  of Double (deep) $Q$-learning can then be written as
\begin{eqnarray}\label{eq:Y_update3}
\overline{Y}^Q_n = r + \gamma\, Q(s',\arg\max_{a\in \mathcal{A}}Q(s',a;\theta^{(n)});\eta^{(n)}).
\end{eqnarray}
Notice that the selection of the action, in the argmax, is
still due to the online weights $\theta^{(n)}$. This means that, as in $Q$-learning, we are still estimating the value of the greedy policy according to the current values, as defined by $\theta^{(n)}$. However, we use the second set of weights $\eta^{(n)}$ to fairly compute the value of this policy. This second set of weights can be updated symmetrically by switching the roles of $\theta$ and $\eta$.

\paragraph{Convergence Guarantee.} 
For DQN, \cite{fan2020theoretical} characterized the approximation error of the $Q$-function by the sum of a statistical error and an algorithmic error, and the latter decays to zero at a geometric rate as the algorithm proceeds. The statistical error characterizes the bias and variance arising from the $Q$-function approximation using the neural network. \cite{cai2019neural} parametrized the $Q$-function by a two-layer neural network and provided a mean-squared sample complexity with sublinear convergence rate for neural TD learning.
The two-layer network in \cite{cai2019neural} with width $m$ is given by
\begin{equation}\label{eqn:two_layer_NN}
    F\big((s,a);\pmb{W}\big)=\frac{1}{\sqrt{m}}\sum_{l=1}^{m} c_l\cdot \sigma\big(\pmb{W}_l^\top (s,a) \big),
\end{equation}
where the activation function $\sigma(\cdot)$ is set to be the ReLU function, and the parameter $\pmb{c}=(c_1,\ldots,c_m)$ is fixed at the initial parameter during the training process and only the weights $\pmb{W}$ are updated. \cite{xu2020finite} considered a more
challenging setting than \cite{cai2019neural}, where the input data are non-i.i.d and the neural network has multiple (more than two) layers and obtained the same sublinear convergence rate. Furthermore, \cite{cayci2021sample} also employed the two-layer neural network in \eqref{eqn:two_layer_NN} and proved that two algorithms used in practice, the projection-free and max-norm regularized \cite{goodfellow2013maxout} neural TD, achieve mean-squared sample complexity of $\widetilde{\mathcal{O}}^\prime(1/\varepsilon^6)$ and $\widetilde{\mathcal{O}}^\prime(1/\varepsilon^4)$, respectively.

\subsection{Deep Policy-based RL Algorithms}\label{subsec:deep_policy_based}

In this section we focus on deep policy-based methods, which are extensions of policy-based methods using neural network approximations. We parameterize the policy $\pi$ by a neural network $F$ with parameter $\theta=(\pmb{W},
\pmb{b})$, that is, $a\sim\pi(s,a;\theta)=f(F((s,a);\theta))$ for some function $f$. A popular choice of $f$ is given by
\[
f(F((s,a);\theta))=\frac{\exp(\tau F((s,a);\theta))}{\sum_{a^\prime\in\mathcal{A}}\exp(\tau F((s,a^\prime);\theta))},
\]
for some parameter $\tau$, which gives an \emph{energy-based} policy (see, e.g.  \cite{haarnoja2017reinforcement,wang2019neural}). 
The policy parameter $\theta$ is updated using the gradient ascent rule given by
\[
\theta^{(n+1)} = \theta^{(n)} + \beta \widehat{\nabla_{\theta}J(\theta^{(n)})}, \qquad n=0,\ldots,N-1,
\]
where $\widehat{\nabla_{\theta}J(\theta^{(n)})}$ is an estimate of the policy gradient.

Using neural networks to parametrize the policy and/or value functions in the vanilla version of policy-based methods discussed in Section \ref{subsec:policy_based_methods} leads to neural Actor-Critic algorithms \cite{wang2019neural}, neural PPO/TRPO \cite{liu2019neural}, and deep DPG (DDPG) \cite{lillicrap2015continuous}. 
In addition, since introducing an entropy term in the objective function encourages policy exploration \cite{haarnoja2017reinforcement} and speeds the learning process \cite{haarnoja2018soft,mei2020} (as discussed in Section \ref{subsec:regularization}), there have been some recent developments in (off-policy) \emph{soft} Actor-Critic algorithms \cite{haarnoja2018soft,haarnoja2018softappl} using neural networks, which solve the RL problem with entropy regularization. Below we introduce the DDPG algorithm, which is one of the most popular deep policy-based methods, and which has been applied in many financial problems.

\paragraph{DDPG.} DDPG is a model-free off-policy Actor-Critic algorithm, first introduced in \cite{lillicrap2015continuous}, which combines the DQN and DPG algorithms. Since its structure is more complex than DQN and DPG, we provide the pseudocode for DDPG in Algorithm \ref{alg:DDPG}. DDPG extends the DQN to continuous action spaces by incorporating DPG to learn a deterministic strategy. To encourage exploration, DDPG uses the following action
\[
a_t \sim \pi^D(s_t;\theta_t) + \epsilon,
\]
where $\pi^D$ is a deterministic policy and $\epsilon$ is a random variable sampled from some distribution $\mathcal{N}$, which can be chosen according to the environment. Note that the algorithm requires a small learning rate $\bar{\beta}\ll 1$ (see line 14 in Algorithm \ref{alg:DDPG}) to improve the stability of learning the target networks. In the same way as DQN (see Section \ref{subsec:deep_value_based}), the DDPG algorithm also uses a replay buffer to improve the performance of neural networks.

\begin{algorithm}[H]
\caption{\textbf{The DDPG Algorithm}}
\label{alg:DDPG}
\begin{algorithmic}[1]
    \STATE \textbf{Input}: an actor $\pi^D(s;\theta)$, a critic network $Q(s,a;\phi)$, learning rates $\beta$ and $\bar{\beta}$, initial parameters $\theta^{(0)}$ and $\phi^{(0)}$
    \STATE Initialize target networks parameters $\bar{\phi}^{(0)}\leftarrow \phi^{(0)}$ and $\bar{\theta}^{(0)}\leftarrow \theta^{(0)}$, 
    \STATE Initialize replay buffer $\mathcal{B}$  
    \FOR{$n=0,\ldots,N-1$}
    \STATE Initialize state $s_1$
    \FOR{$t=0,\ldots,M-1$} 
    \STATE Select action  $a_t\sim\pi^D(s_t;\theta^{(n)})+\epsilon_t$  with $\epsilon_t\sim\mathcal{N}$
    \STATE Execute action $a_t$ and observe reward $r_t$ and observe new state $s_{t+1}$
    \STATE Store transition $(s_t,a_t,r_t,s_{t+1})$ in $\mathcal{B}$
     \IF{$|\mathcal{B}|>B$}
    \STATE Sample a random mini-batch of $B$ transitions $\{(s_{(i)},a_{(i)},r_{(i)},s_{(i+1)})\}_{i=1}^B$ from $\mathcal{B}$
    \STATE Set the target $Y_i = r_i+\gamma Q(s_{i+1},\pi^D(s_{i+1};\bar{\theta}^{(n)});\bar{\phi}^{(n)})$.
    \STATE Update the critic by minimizing the loss: $\phi^{(n+1)}=\phi^{(n)}-\beta\nabla_{\phi}\mathcal{L}_{\rm DDPG}(\phi^{(n)})$ with 
    $\mathcal{L}_{\rm DDPG}(\phi^{(n)})=\frac{1}{B}\sum_{i=1}^B (Y_i-Q(s_i,a_i;\phi^{(n)}))^2$.
    \STATE Update the actor by using the sampled policy gradient: $ \theta^{(n+1)}=\theta^{(n)}+\beta\nabla_{\theta}J(\theta^{(n)})$ with 
    \[
    \nabla_{\theta}J(\theta^{(n)})\approx \frac{1}{B}\sum_{i=1}^B\nabla_a Q(s,a;\phi^{(n)})|_{s=s_i,a=a_i}\nabla_{\theta}\pi^D(s;\theta^{(n)})|_{s=s_i}.
    \]
    \STATE Update the target networks: 
    \begin{eqnarray*}
    \bar{\phi}^{(n+1)}&\leftarrow& \bar{\beta} \phi^{(n+1)} + (1-\bar{\beta}) \bar{\phi}^{(n)}\\
    \bar{\theta}^{(n+1)}&\leftarrow& \bar{\beta} \theta^{(n+1)} + (1-\bar{\beta}) \bar{\theta}^{(n)}
    \end{eqnarray*}
   \ENDIF
    \ENDFOR
    \ENDFOR
\end{algorithmic}
\end{algorithm}

\paragraph{Convergence Guarantee.} By parameterizing the policy and/or value functions using a two-layer neural network given in \eqref{eqn:two_layer_NN}, \cite{liu2019neural} provided a mean-squared sample complexity for neural PPO and TRPO algorithms with sublinear convergence rate; \cite{wang2019neural} studied neural Actor-Critic methods where the actor updates using (1) vanilla policy gradient or (2) natural policy gradient, and in both cases the critic updates using TD(0). They proved that in case (1) the algorithm converges to a stationary point at a sublinear rate and they also established the global optimality of all stationary points under mild
regularity conditions. In case (2) the algorithm was proved to achieve a mean-squared sample complexity with sublinear convergence rate. To the best of our knowledge, no convergence guarantee has been established for the DDPG (and DPG) algorithms.


\section{Applications in Finance}\label{sec:fin_app}

The availability of data from electronic markets and the recent developments in RL together have led to a rapidly growing recent body of work applying RL algorithms to decision-making problems in electronic markets. Examples include optimal execution, portfolio optimization, option pricing and hedging, market making, order routing, as well as robot advising.

In this section, we start with a brief overview of electronic markets and some discussion of market microstructure in Section \ref{sec:electronic_market}. We then introduce several applications of RL in finance. In particular, optimal execution for a single asset is introduced in Section \ref{sec:optimal_execution} and portfolio optimization problems across multiple assets  is discussed in Section  \ref{sec:portfolio_optimization}. This is followed by sections on option pricing, robo-advising, and smart order routing. In each we introduce the underlying problem and basic model before looking at recent RL approaches used in tackling them.

It is worth noting that there are some open source projects that provide full pipelines for implementing different RL algorithms in financial applications \cite{liu2021finrl,liu2022finrl}.

\subsection{Preliminary: Electronic Markets and  Market Microstructure}\label{sec:electronic_market}

Many recent decision-making problems in finance are centered around electronic markets. We give a brief overview of this type of market and discuss two popular examples -- central limit order books and electronic over-the-counter markets. For more in-depth discussion of electronic markets and market microstructure, we refer  the reader to  the books \cite{cartea2015algorithmic} and \cite{lehalle2018market}.

\paragraph{Electronic Markets.} Electronic markets have emerged as popular venues for the trading of a wide variety of financial assets. Stock exchanges in many countries, including Canada, Germany, Israel, and the United Kingdom, have adopted electronic platforms to trade equities, as has Euronext, the market combining several former European stock exchanges. In the United States, electronic communications networks (ECNs) such as Island, Instinet, and Archipelago (now ArcaEx) use an electronic order book structure to trade as much as 45\% of the volume on the NASDAQ stock market. Several electronic systems trade corporate bonds and government bonds (e.g., BrokerTec, MTS), while in foreign exchange, electronic systems such as EBS and Reuters dominate the
trading of currencies. Eurex, the electronic Swiss-German exchange, is now the world's largest futures market, while options have been traded in electronic markets since the opening of the International Securities Exchange in 2000. Many such electronic markets are organized as electronic Limit Order Books (LOBs). In this structure, there is no designated liquidity provider such as a specialist or a dealer. Instead, liquidity arises endogenously from the submitted orders of traders. Traders who submit orders to buy or sell the asset at a particular price are said to ``make''
liquidity, while traders who choose to hit existing orders are said to ``take'' liquidity. Another major type of electronic market without a central LOB is the Over-the-Counter (OTC) market where the orders are executed directly between two parties, the dealer and the client, without the supervision of an exchange. Examples include BrokerTec and MTS for trading corporate bonds and government bonds.

\begin{figure}[H]
  \centering
    \includegraphics[width=0.55\textwidth]{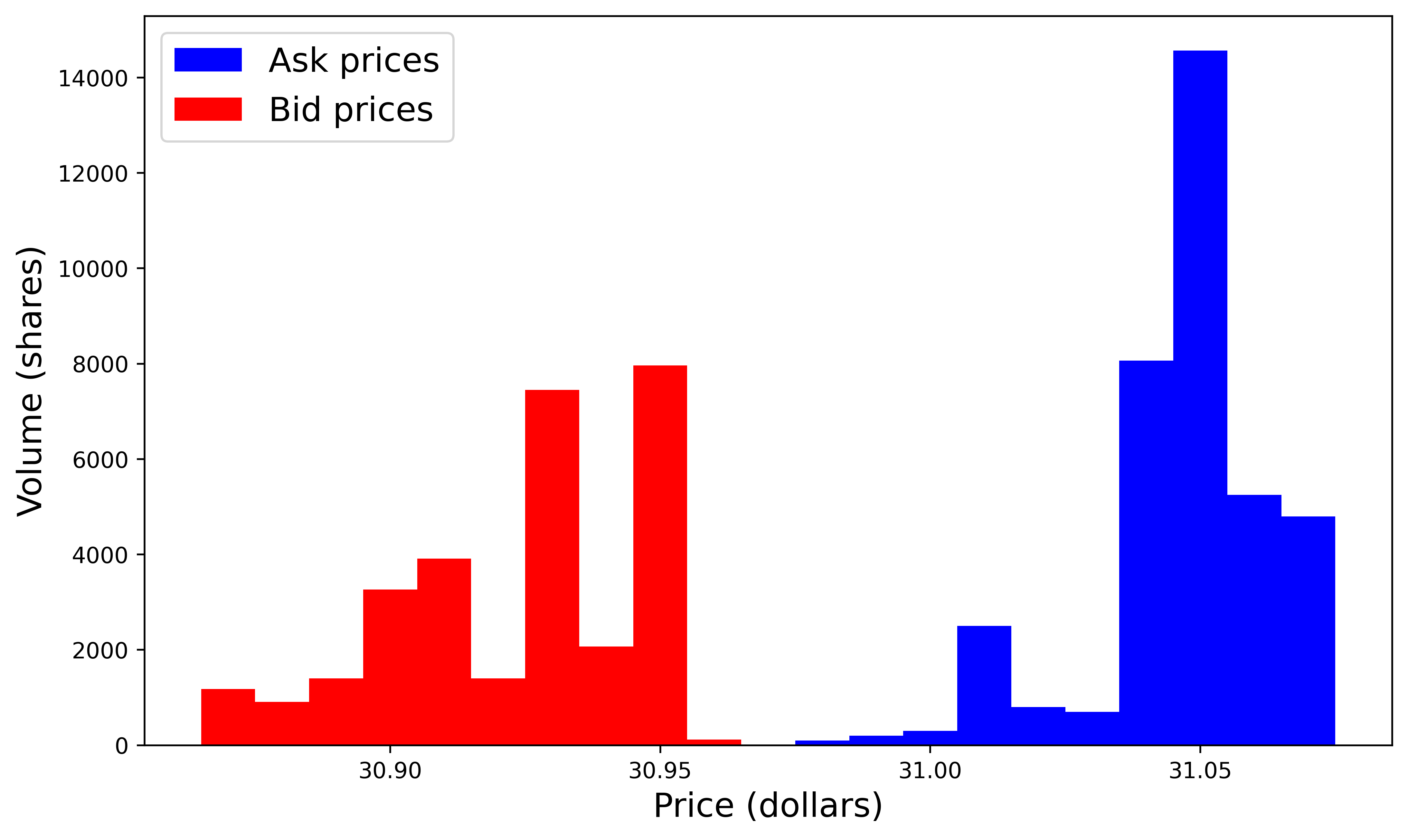}
  \caption{\label{fig:LOB_configuration}A snapshot of the LOB of MSFT(Microsoft) stock at 9:30:08.026 am on 21 June, 2012 with ten levels of bid/ask prices. 
  }
\end{figure}

\paragraph{Limit Order Books.} An LOB is a list of orders that a trading venue, for example the NASDAQ exchange, uses to record the interest of buyers and sellers in a particular financial asset or instrument. There are two types of orders the buyers (sellers) can submit: a limit buy (sell) order with a preferred price for a given volume of the asset or a market buy (sell) order with a given volume which will be immediately executed with the best available limit sell (buy) orders. Therefore limit orders have a price guarantee but are not guaranteed to be executed, whereas market orders are executed immediately at the best available price. The lowest price of all sell limit orders is called the \emph{ask} price, and the highest price of all buy limit orders is called the \emph{bid} price. The difference between the ask and bid is known as the \emph{spread} and the average of the ask and bid is known as the \emph{mid-price}. A snapshot of an LOB\footnote{We use NASDAQ ITCH data taken from Lobster https://lobsterdata.com/.} with 10 levels of bid/ask prices is shown in Figure \ref{fig:LOB_configuration}.     

A matching engine is used to match the incoming buy and sell orders. This typically follows the price-time priority rule \cite{preis2011price}, whereby orders are first ranked according to their price. Multiple orders having the same price are then ranked according to the time they were entered. If the price and time are the same for the incoming orders, then the larger order gets executed first. The matching engine uses the LOB to store pending orders that could not be executed upon arrival.

\paragraph{Over-the-counter Markets.} OTC or off-exchange trading is done directly between two parties, without the supervision of an exchange. Many OTC markets organized around dealers, including corporate bond markets in many countries, have also undergone upheaval linked to electronification over the last ten years. The electronification process is dominated by Multi-dealer-to-client (MD2C) platforms enabling clients to send the same request for a quote (RFQ) to several dealers simultaneously and therefore instantly put the dealers into competition with one another. These dealers can then each provide a price to the client for the transaction (not necessarily
the price the dealer has streamed). Dealers know the identity of the client (which differs from most of the systems organized around a central LOB) and the
number of requested dealer prices. However,
they do not see the prices that are streamed by the other dealers. They only see a composite price at the bid and offer, based on some of the best streamed prices.
The client progressively receives the answers to the RFQ and can deal at any time
with the dealer who has proposed the best price, or decide not to trade. Each dealer knows whether a deal was done (with her/him, but also with another
dealer - without knowing the identity of this dealer) or not. If a transaction occurred, the best dealer usually knows the cover price (the second best bid price in the RFQ), if there is one. We refer the reader to \cite{fermanian2015agents} for a more in-depth discussion of MD2C bond trading platforms.

\paragraph{Market Participants.}
When considering different market participants it is sometimes helpful to classify them based on their objectives and trading strategies. Three primary classes are the following \cite{cartea2015algorithmic}:
\begin{itemize}
    \item Fundamental (or noise or liquidity) traders: those who are driven by economic fundamentals outside the exchange.
    \item Informed traders: traders who profit from leveraging information not reflected in market prices by trading assets in anticipation of their appreciation or depreciation.
    \item Market makers: professional traders who profit from facilitating exchange in a particular asset and exploit their skills in executing trades.
\end{itemize}
The effects of the interactions amongst all these traders is one of the key issues studied in the field of market microstructure. How to improve  trading decisions from the perspective of one class of market participants, while strategically interacting with the others, is one of the big challenges in the field. The recent literature has seen many attempts to exploit RL techniques to tackle these problems.



\subsection{Optimal Execution}\label{sec:optimal_execution}
Optimal execution is a fundamental problem in financial modeling. The simplest version is the case of a trader who wishes to buy or sell a given amount of a single asset within a given time period. The trader seeks strategies that maximize their return from, or alternatively, minimize the cost of, the execution of the transaction. 

\paragraph{The Almgren--Chriss Model.} A classical framework for optimal execution is due to Almgren--Chriss \cite{AC2001}. In this setup a trader is required to sell an amount $q_0$ of an asset, with
price $S_0$ at time 0, over the time period $[0,T]$ with trading decisions made at discrete time points
$t=1,\ldots,T$. The final inventory $q_T$ is required to be zero. Therefore the goal is to determine the liquidation strategy $u_1,u_2,\ldots,u_{T}$, where $u_t$ ($t=1,2,\ldots,T$) denotes the amount of the asset to sell at time $t$. It is assumed that selling quantities of the asset will have two types of price impact -- a temporary impact which refers to any temporary price movement due to the supply-demand imbalance caused by the selling, and a permanent impact, which is a long-term effect on the `equilibrium' price due to the trading, that will remain at least for the trading period. 

We write $S_t$ for the asset price at time $t$. The Almgren--Chriss model assumes that the asset price evolves according to the discrete arithmetic random walk 
\[
S_t = S_{t-1}+\sigma\xi_{t}-g(u_t),\qquad t=1,2,\ldots,T
\]
where $\sigma$ is the (constant) volatility parameter, $\xi_t$ are independent random variables drawn from a distribution with zero mean and unit variance, and $g$ is a function of the trading strategy $u_t$ that measures the permanent impact. The inventory process $\{q_t\}_{t=0}^T$ records the current holding
in the asset, $q_t$, at each time $t$. Thus we have
\[
q_t = q_{t-1}-u_t. 
\]
Selling the asset may cause a temporary price impact, that is a drop in the average price per share, thus the actual price per share received is 
\[
\widetilde{S}_t = S_{t-1}-h(u_t),
\]
where $h$ is a function which quantifies this temporary price impact. The cost of this trading trajectory is defined to be the difference between the initial book value and the revenue, given by $C = q_0S_0 - \sum_{t=1}^T q_t\widetilde{S}_t$ with mean and variance
\[
 \mathbb{E}[C] = \sum_{t=1}^T \big(q_tg(u_t)+u_th(u_t)\big),\qquad {\rm Var}(C)=\sigma^2\sum_{t=1}^Tq_t^2.
\]
A trader thus would like to minimize her expected cost as well as the variance of the cost, which gives the optimal execution problem in this model as 
\[
\min_{\{u_t\}_{t=1}^{T}}\big(\mathbb{E}[C]+\lambda {\rm Var}(C)\big),
\]
where $\lambda\in\mathbb{R}$ is a measure of risk-aversion.

When we assume that both the price impacts are linear, that is 
\[
g(u_t) = \gamma u_t, \qquad h(u_t) = \eta u_t,
\]
where $\gamma$ and $\eta$ are the permanent and temporary price impact parameters, \cite{AC2001} derive the general solution for the Almgren--Chriss model. This is given by
\begin{equation*}
     u_j = \frac{2\,{\rm sinh}\big(\frac{1}{2}\kappa \big)}{{\rm sinh} (\kappa T)}{\rm cosh}\left(\kappa\left(T-t_{j-\frac{1}{2}}\right)\right)q_0, \qquad j=1,2,\ldots,T,
\end{equation*}
with 
\[
\kappa = {\rm cosh}^{-1}\left(\frac{\tilde{\kappa}^2}{2}+1\right), \qquad \tilde{\kappa}^2 = \frac{\lambda\sigma^2}{\eta(1-\frac{\gamma}{2\eta})}.
\]
The corresponding optimal inventory trajectory is 
\begin{equation}\label{eq:amgren-chriss}
    q_j = \frac{{\rm sinh}\big(\kappa (T-t_j)\big)}{{\rm sinh} (\kappa T)}q_0, \qquad j=0,1,\ldots,T.
\end{equation}
The above Almgren--Chriss framework for liquidating a single asset can be extended to the case of multiple assets \cite[Appendix A]{AC2001}. We also note that the general solution to the Almgren-Chriss model had been constructed previously in \cite[Chapter 16]{grinold2000active}.

This simple version of the Almgren--Chriss framework has a closed-form solution but it relies heavily on the assumptions of the dynamics and the linear form of the permanent and temporary price impact. The mis-specification of the dynamics and market impacts may lead to undesirable  strategies and potential losses. In addition, the solution in \eqref{eq:amgren-chriss} is a pre-planned strategy that does not depend on real-time market conditions.  Hence this strategy may miss certain opportunities when the market moves. This motivates the use of an RL approach which is more flexible and able to incorporate market conditions when making decisions. 

\paragraph{Evaluation Criteria and Benchmark Algorithms.} Before discussing the RL approach, we introduce several widely-used criteria to evaluate the performance of execution algorithms in the literature such as the Profit and Loss (PnL), Implementation Shortfall, and the Sharp ratio. The PnL is the {\it final} profit or loss induced by a given execution algorithm over the whole time period, which is made up of transactions at all time points.
 The Implementation Shortfall \cite{perold1988implementation} for an execution algorithm is defined as the difference between the {PnL} of the algorithm and the {PnL} received by trading the entire amount of the asset instantly. The Sharpe ratio \cite{sharpe1966} is defined as the ratio of expected return to standard deviation of the return; thus it measures return per unit of risk. Two popular variants of the Sharpe ratio are the differential Sharpe ratio \cite{moody1998dSharpe} and the Sortino ratio \cite{sortino1994}.

In addition, some classical pre-specified strategies are used as benchmarks to evaluate the performance of a given RL-based execution strategy. Popular choices include executions strategies based on Time-Weighted Average Price (TWAP) and Volume-Weighted Average Price (VWAP) as well as the Submit and Leave (SnL) policy where a trader places a sell order for all shares at a fixed limit order price, and goes to the market with any unexecuted shares remaining at time $T$.

\paragraph{RL Approach.} We first provide a brief overview of the existing literature on RL for optimal execution. The most popular types of RL methods that have been used in optimal execution problems are $Q$-learning algorithms and (double) DQN \cite{HW2014,NLJ2018,ZZR2020,JK2019,DGK2019,SHYO2014,NFK2006,cartea2021survey}. Policy-based algorithms are also popular in this field, including (deep) policy gradient methods \cite{hambly2020policy,ZZR2020}, A2C \cite{ZZR2020}, PPO \cite{DGK2019,LB2020}, and DDPG \cite{Ye2020}. The benchmark strategies studied in these papers include the Almgren--Chriss solution \cite{HW2014,hambly2020policy}, the TWAP strategy \cite{NLJ2018,DGK2019,LB2020}, the VWAP strategy \cite{LB2020}, and the SnL policy \cite{NFK2006,Ye2020}. In some models the trader is allowed to buy or sell the asset at each time point \cite{JK2019,ZZR2020,WWMD2019,Deng2016}, whereas there are also many models where only one trading direction is allowed \cite{NFK2006,HW2014,hambly2020policy,NLJ2018,DGK2019,SHYO2014,Ye2020,LB2020}. The state variables are often composed of time stamp, the market attributes including (mid-)price of the asset and/or the spread, the inventory process and past returns. The control variables are typically set to be the amount of asset (using market orders) to trade and/or the relative price level (using limit orders) at each time point. Examples of reward functions include cash inflow or outflow (depending on whether we sell or buy) \cite{SHYO2014,NFK2006}, implementation shortfall \cite{HW2014}, profit \cite{Deng2016}, Sharpe ratio \cite{Deng2016}, return \cite{JK2019}, and PnL \cite{WWMD2019}. Popular choices of performance measure include Implementation Shortfall \cite{HW2014,hambly2020policy}, PnL (with a penalty term of transaction cost) \cite{NLJ2018,WWMD2019,Deng2016}, trading cost \cite{NFK2006,SHYO2014}, profit \cite{Deng2016,JK2019},  Sharpe ratio \cite{Deng2016,ZZR2020}, Sortino ratio \cite{ZZR2020}, and return \cite{ZZR2020}.

We now discuss some more details of the RL algorithms and experimental settings in the above papers. For value-based algorithms, \cite{NFK2006} provided the first large scale empirical analysis of a RL method applied to optimal execution problems. They focused on a modified $Q$-learning algorithm to select price levels for limit order trading, which leads to significant improvements over simpler forms of optimization such as the SnL policies in terms of trading costs. \cite{SHYO2014} proposed a risk-averse RL algorithm for optimal liquidation, which can be viewed as a generalization of \cite{NFK2006}. This algorithm achieves substantially lower trading costs over the period when the 2010 flash-crash happened compared with the risk-neutral RL algorithm in \cite{NFK2006}. \cite{HW2014} combined the Almgren--Chriss solution and the $Q$-learning algorithm and showed that they are able to improve the Implementation Shortfall of the Almgren--Chriss solution by up to 10.3\% on average, using LOB data which includes five price levels.  \cite{NLJ2018} proposed a modified double DQN algorithm for optimal execution and showed that the algorithm outperforms the TWAP strategy on seven out of nine stocks using PnL as the performance measure. They added a single one-second time step $\Delta T$ at the end of the horizon to guarantee that all shares are liquidated over $T+\Delta T$. \cite{JK2019} designed a trading system based on DQN which determines both the trading direction and the number of shares to trade. Their approach was shown to increase the total profits by at least four times in four different index stocks compared with {a benchmark trading model which trades a fixed number of shares each time}. They also used \emph{transfer} learning to avoid overfitting, where some knowledge/information is reused when learning in a related or similar situation. 

For policy-based algorithms,  \cite{Deng2016} combined deep learning with RL to determine whether to sell, hold, or buy at each time point. In the first step of their model, neural networks are used to summarize the market features and in the second step the RL part makes trading decisions. The proposed method was shown to outperform several other deep learning and deep RL models in terms of PnL, total profits, and Sharpe ratio. They suggested that in practice, the Sharpe ratio was more recommended as the reward function compared with total profits. \cite{Ye2020} used the DDPG algorithm for optimal execution over a short time horizon (two minutes) and designed a network to extract features from the market data. Experiments on real LOB data with 10 price levels show that the proposed approach significantly outperforms the existing methods, including the SnL policy (as baseline), the $Q$-learning algorithm, and the method in \cite{HW2014}. \cite{LB2020} proposed an adaptive framework based on PPO with neural networks including LSTM and fully-connected networks, and showed that the framework outperforms the baseline models including TWAP and VWAP, as well as several deep RL models on most of 14 US equities. \cite{hambly2020policy} applied the (vanilla) policy gradient method to the LOB data of five stocks in different sectors and showed that they improve the Implementation Shortfall of the Almgren--Chriss solution by around 20\%. \cite{leal2020learning} used  neural networks to learn the mapping between the risk-aversion parameter and the optimal control  with potential market impacts incorporated.

{For comparison between value-based and policy-based algorithms,} \cite{DGK2019} explored double DQN and PPO algorithms in different market environments -- when the benchmark TWAP is optimal, PPO is shown to converge to TWAP whereas double DQN may not; when TWAP is not optimal, both algorithms outperform this benchmark. \cite{ZZR2020} showed that DQN, policy gradient, and A2C outperform several baseline models including classical time-series momentum strategies, on test data of 50 liquid futures contracts. Both continuous and discrete action spaces were considered in their work. They observed that DQN achieves the best performance and the second best is the A2C approach.

{In addition, model-based RL algorithms have also been used for optimal execution.} \cite{WWMD2019} built a profitable electronic trading agent to place buy and sell orders using model-based RL, which outperforms two benchmarks strategies in terms of PnL on LOB data. They used a recurrent neural network to learn the state transition probability. We note that multi-agent RL has also been used to address the optimal execution problem, see for example \cite{BL2019,KFMW2020}.

\subsection{Portfolio Optimization}\label{sec:portfolio_optimization}

In portfolio optimization problems, a trader needs to select and trade the best portfolio of assets in order to maximize some objective function, which typically includes the expected return and some measure of the risk. The benefit of investing in such portfolios is that the diversification of investments achieves higher return per unit of risk than only investing in a single asset (see, e.g. \cite{zivot2017introduction}).

\paragraph{Mean-Variance Portfolio Optimization.}
The first significant mathematical model for portfolio optimization is the \emph{Markowitz} model \cite{Markowitz1952}, also called the \emph{mean-variance} model, where an investor seeks a portfolio to maximize the expected total return for any given level of risk measured by variance. This is a single-period optimization problem and is then generalized to multi-period portfolio optimization problems in \cite{mossin1968optimal,hakansson1971multi,samuelson1975lifetime,merton1974,steinbach1999markowitz,li2000multiperiodMV}. In this mean-variance framework, the risk of a portfolio is quantified by the variance of the wealth and the optimal investment strategy is then sought to maximize the final wealth penalized by a variance term. The mean-variance framework is of particular interest because it not only captures both the portfolio return and the risk, but also suffers from the 
\emph{time-inconsistency} problem \cite{strotz1955myopia,vigna2016time,xiao2020corre}. That is the optimal strategy selected at time $t$ is no longer optimal at time $s>t$ and the Bellman equation does not hold. A breakthrough was made in \cite{li2000multiperiodMV}, in which they were the first to derive the analytical solution to the discrete-time multi-period mean-variance problem. They applied an \emph{embedding} approach, which transforms the mean-variance problem to an LQ problem where classical approaches can be used to find the solution. The same approach was then used to solve the continuous-time mean-variance problem \cite{zhou2000continuous}. In addition to the embedding scheme, other methods including the consistent planning approach \cite{basak2010dynamic,bjork2010general} and the dynamically optimal strategy \cite{pedersen2017optimal} have also been applied to solve the time-inconsistency problem arising in the mean-variance formulation of portfolio optimization.

Here we introduce the multi-period mean-variance portfolio optimization problem as given in \cite{li2000multiperiodMV}. Suppose there are $n$ risky assets in the market and an investor enters the market at time 0 with initial wealth $x_0$. The goal of the investor is to reallocate his wealth at each time point $t=0,1,\ldots,T$ among the $n$ assets to achieve the optimal trade off between the return and the risk of the investment. The random rates of return of the assets at $t$ is denoted by $\pmb{e}_t =(e_t^1,\ldots,e_t^n)^\top$, where $e_t^i$ ($i=1,\ldots,n$) is the rate of return of the $i$-th asset at time $t$. The vectors $\pmb{e}_t$,  $t=0,1,\ldots,T-1$, are assumed to be statistically independent (this independence assumption can be relaxed, see, e.g. \cite{xiao2020corre}) with known mean  $\pmb{\mu}_t=(\mu_t^1,\ldots,\mu_t^n)^\top\in\mathbb{R}^{n}$ and known standard deviation $\sigma_{t}^i$ for $i=1,\ldots,n$ and $t=0,\ldots,T-1$. The covariance matrix is denoted by $\pmb{\Sigma}_t\in\mathbb{R}^{n\times n}$, where $[\pmb{\Sigma}_{t}]_{ii}=(\sigma_{t}^i)^2$ and $[\pmb{\Sigma}_t]_{ij}=\rho_{t}^{ij}\sigma_{t}^i\sigma_{t}^j$ for $i,j=1,\ldots,n$ and $i\neq j$, where $\rho_{t}^{ij}$ is the correlation between assets $i$ and $j$ at time $t$. We write $x_t$ for the wealth of the investor at time $t$ and $u_t^i$ ($i=1,\ldots,n-1$) for the amount invested in the $i$-th asset at time $t$. Thus the amount invested in the $n$-th asset is $x_t-\sum_{i=1}^{n-1} u_t^i$. An investment strategy is denoted by $\pmb{u}_t= (u_t^1,u_t^2,\ldots,u_t^{n-1})^\top$ for $t=0,1,\ldots,T-1$, and the goal is to find an optimal strategy such that the portfolio return is maximized while minimizing the risk of the investment, that is,
\begin{eqnarray}\label{eqn:MV_obj}
\max_{\{\pmb{u}_t\}_{t=0}^{T-1}}\mathbb{E}[x_T] - \phi {\rm Var}(x_T),
\end{eqnarray}
subject to 
\begin{equation}\label{eqn:MV_state}
    x_{t+1} = \sum_{i=1}^{n-1} e_t^i u_t^i + \left(x_t-\sum_{i=1}^{n-1} u_t^i\right)e_t^n,\qquad t=0,1,\ldots,T-1,
\end{equation}
where $\phi$ is a weighting parameter balancing risk, represented by the variance of $x_T$, and return. As mentioned above, this framework is embedded into an LQ problem in \cite{li2000multiperiodMV}, and the solution to the LQ problem gives the solution to the above problem \eqref{eqn:MV_obj}-\eqref{eqn:MV_state}. The analytical solution derived in \cite{li2000multiperiodMV} is of the following form
\[
\pmb{u}_t^* (x_t) = \alpha_t x_t + \beta_t,
\]
where $\alpha_t$ and $\beta_t$ are explicit functions of $\pmb{\mu}_t$ and $\pmb{\Sigma}_t$, which are omitted here and can be found in \cite{li2000multiperiodMV}.

The above framework has been extended in different ways, for example, the risk-free asset can also be involved in the portfolio, and one can maximize the cumulative form of \eqref{eqn:MV_obj} rather than only focusing on the final wealth $x_T$. For more details about these variants and solutions, see \cite{xiao2020corre}. In addition to the mean-variance framework, other major paradigms in portfolio optimization are the Kelly Criterion and Risk Parity. We refer to \cite{Sato2019Survey} for a review of these optimal control frameworks and popular model-free RL algorithms for portfolio optimization.

Note that the classical stochastic control approach for portfolio optimization problems across multiple  assets requires both a realistic representation of the temporal dynamics  of individual assets, as well as an adequate representation  of their co-movements. This is extremely difficult when the assets belong to different classes (for example, stocks, options, futures, interest rates and their derivatives). On the other hand, the model-free RL approach does not rely on the specification of the joint dynamics across assets. 

\paragraph{RL Approach.} Both value-based methods such as $Q$-learning \cite{du2016algorithm,pendharkar2018trading}, SARSA \cite{pendharkar2018trading}, and DQN \cite{PSC2020}, and policy-based algorithms such as DPG and DDPG \cite{xiong2018practical,JXL2017,YLKSD2019,liang2018adversarial,aboussalah2020value,cong2021alphaportfolio} have been applied to solve portfolio optimization problems. The state variables are often composed of time, asset prices, asset past returns, current holdings of assets, and remaining balance. The control variables are typically set to be the amount/proportion of wealth invested in each component of the portfolio. Examples of reward signals include  portfolio return  \cite{JXL2017,pendharkar2018trading,YLKSD2019}, (differential) Sharpe ratio \cite{du2016algorithm,pendharkar2018trading}, and profit \cite{du2016algorithm}. The benchmark strategies include Constantly Rebalanced Portfolio (CRP) \cite{YLKSD2019,JXL2017} where at each period the portfolio is rebalanced to the initial wealth distribution among the assets, and the buy-and-hold or do-nothing strategy \cite{PSC2020,aboussalah2020value} which does not take any action but rather holds the initial portfolio until the end. The performance measures studied in these papers include the Sharpe ratio \cite{YLKSD2019,WZ2019,xiong2018practical,JXL2017,liang2018adversarial,PSC2020,wang2019}, the Sortino ratio \cite{YLKSD2019}, portfolio returns \cite{wang2019,xiong2018practical,liang2018adversarial,PSC2020,YLKSD2019,aboussalah2020value}, portfolio values \cite{JXL2017,xiong2018practical,pendharkar2018trading}, and cumulative profits \cite{du2016algorithm}. Some models incorporate the transaction costs \cite{JXL2017,liang2018adversarial,du2016algorithm,PSC2020,YLKSD2019,aboussalah2020value} and investments in the risk-free asset \cite{YLKSD2019,WZ2019,wang2019,du2016algorithm,JXL2017}.

For value-based algorithms, 
\cite{du2016algorithm} considered the portfolio optimization problems of a risky asset and a risk-free asset. They compared the performance of the $Q$-learning algorithm and a Recurrent RL (RRL) algorithm under three different value functions including the Sharpe ratio, differential Sharpe ratio, and profit. The RRL algorithm is a policy-based method which uses the last action as an input. They concluded that the $Q$-learning algorithm is more sensitive to the choice of value function and has less stable performance than the RRL algorithm. They also suggested that the (differential) Sharpe ratio is preferred rather than the profit as the reward function. \cite{pendharkar2018trading} studied a two-asset personal retirement  portfolio optimization problem, in which traders who manage retirement funds are restricted from making frequent trades and they can only access limited information. They tested the performance of three algorithms: SARSA and $Q$-learning methods with discrete state and action space that maximize either the portfolio return or differential Sharpe ratio, and a TD learning method with discrete state space and continuous action space that maximizes the portfolio return. The TD method learns the portfolio returns by using a linear regression model and was shown to outperform the other two methods in terms of portfolio values. \cite{PSC2020} proposed a portfolio trading strategy based on DQN which chooses to either hold, buy or sell a pre-specified quantity of the asset at each time point. In their experiments with two different three-asset portfolios, their trading strategy is superior to four benchmark strategies including the do-nothing strategy and a random strategy (take an action in the feasible space randomly) using performance measures including the cumulative return and the Sharpe ratio. \cite{dixon2020g} applies a G-learning-based algorithm, a probabilistic extension of $Q$-learning
which scales to high dimensional portfolios while providing a flexible choice of utility functions, for wealth management problems. In addition, the authors also  extend the G-learning algorithm
 to the setting of Inverse Reinforcement Learning (IRL) where rewards collected by the agent are not observed, and should instead be inferred.

For policy-based algorithms, \cite{JXL2017} proposed a framework combining neural networks with DPG. They used a so-called Ensemble of Identical Independent Evaluators (EIIE) topology to predict the potential growth of the assets in the immediate future using historical data which includes the highest, lowest, and closing prices of portfolio components. The experiments using real cryptocurrency market data showed that their framework achieves higher Sharpe ratio and cumulative portfolio values compared with three benchmarks including CRP and several published RL models. \cite{xiong2018practical} explored the DDPG algorithm for the portfolio selection of 30 stocks, where at each time point, the agent can choose to buy, sell, or hold each stock. The DDPG algorithm was shown to outperform two classical strategies including the min-variance portfolio allocation method \cite{yang2018practical} in terms of several performance measures including final portfolio values, annualized return, and Sharpe ratio, using historical daily prices of the 30 stocks. \cite{liang2018adversarial} considered the DDPG, PPO, and policy gradient method with an adversarial learning scheme which learns the execution strategy using noisy market data. They applied the algorithms for portfolio optimization to five stocks using market data and the policy gradient method with adversarial learning scheme achieves higher daily return and Sharpe ratio than the benchmark CRP strategy. They also showed that the DDPG and PPO algorithms fail to find the optimal policy even in the training set. \cite{YLKSD2019} proposed a model using DDPG, which includes prediction of the assets future price movements based on historical prices and synthetic market data generation using a \emph{Generative Adversarial Network} (GAN) \cite{goodfellow2014GAN}. The model outperforms the benchmark CRP and the model considered in \cite{JXL2017} in terms of several performance measures including Sharpe ratio and Sortino ratio. \cite{cong2021alphaportfolio} embedded alpha portfolio strategies into a deep policy-based method and designed a framework which is easier to interpret.
Using Actor-Critic methods, \cite{aboussalah2020value} combined the mean-variance framework (the actor determines the policy using the mean-variance framework)  and the Kelly Criterion framework (the critic evaluates the policy using their growth rate). They studied eight policy-based algorithms including DPG, DDPG, and PPO, among which DPG was shown to achieve the best performance.

In addition to the above discrete-time models, \cite{WZ2019} studied the entropy-regularized continuous-time mean-variance framework with one risky asset and one risk-free asset. They proved that the optimal policy is Gaussian with decaying variance and proposed an Exploratory Mean-Variance (EMV) algorithm, which consists of three procedures: policy evaluation, policy improvement, and a self-correcting scheme for learning the Lagrange multiplier. They showed that this EMV algorithm outperforms two benchmarks, including the analytical solution with estimated model parameters obtained from the classical maximum likelihood method \cite[Section 9.3.2]{campbell2012econometrics} and a DDPG algorithm, in terms of annualized sample mean and sample variance of the terminal wealth, and Sharpe ratio in their simulations. The continuous-time framework in \cite{WZ2019} was then generalized in \cite{wang2019} to large scale portfolio selection setting with $d$ risky assets and one risk-free asset. The optimal policy is shown to be multivariate Gaussian with decaying variance. They tested the performance of the generalized EMV algorithm on price data from the stocks in the S\&P 500 with $d\geq 20$ and it outperforms several algorithms including DDPG.

\subsection{Option Pricing and Hedging}

Understanding how to price and hedge financial derivatives is a cornerstone of modern mathematical and computational finance due to its importance in the finance industry.  A financial derivative is a contract that derives its value from the performance of an underlying entity. For example, a call or put \emph{option} is a contract which gives the holder the right, but not the obligation, to buy or sell an underlying asset or instrument at a specified strike price prior to or on a specified date called the expiration date. Examples of option types include European options which can only be exercised at expiry, and American options which can be exercised at any time before the option expires. 

\paragraph{The Black-Scholes Model.} One of the most important mathematical models for option pricing is the Black-Scholes or Black-Scholes-Merton (BSM) model \cite{BS1973,merton1973}, in which we aim to find the price of a European option $V(S_t,t)$ ($0\leq t\leq T$) with underlying stock price $S_t$, expiration time $T$, 
and payoff at expiry $P(S_T)$. The underlying stock price $S_t$ is assumed to be non-dividend paying and to follow a geometric Brownian motion
\[
d S_t = \mu S_t dt+\sigma S_tdW_t,
\]
where $\mu$ and $\sigma$ are called the drift and volatility parameters of the underlying asset, and $W = \{W_t\}_{0\leq t \leq T}$ is a standard Brownian motion defined on a filtered probability space $(\Omega,\mathcal{F},\{\mathcal{F}\}_{0\leq t \leq T},\mathbb{P})$. 
The key idea of the BSM is that European options can be perfectly replicated by a continuously rebalanced portfolio of the underlying asset and a riskless asset, under the assumption that there are no market frictions, and that trading can take place continuously and in arbitrarily small quantities.
If the derivative can be replicated, by analysing the cost of the hedging strategy, the derivative's price must satisfy the following Black-Scholes partial differential equation
\begin{equation}\label{eqn:BS_equation}
    \frac{\partial V}{\partial t}(s,t)+\frac{1}{2}\sigma^2s^2\frac{\partial^2 V}{\partial s^2}(s,t)+rs\,\frac{\partial V}{\partial s}(s,t) -rV(s,t)=0,
\end{equation}
with terminal condition $V(s,T)=P(s)$ and where $r$ is the known (constant) risk-free interest rate. 
When we have a \emph{call} option with payoff $P(s)=\max(s-K,0)$, giving the option buyer the right to buy the underlying asset, the solution to the Black-Scholes equation is given by
\begin{equation*}
    V(s,t) = N(d_1) s - N(d_2)Ke^{-r(T-t)},
\end{equation*}
where $N(\cdot)$ is the standard normal cumulative distribution function, and $d_1$ and $d_2$ are given by
\[
d_1=\frac{1}{\sigma\sqrt{T-t}}\left(\ln\left(\frac{s}{K}\right)+\left(r+\frac{\sigma^2}{2}\right)(T-t)\right),\qquad d_2=d_1-\sigma\sqrt{T-t}.
\]
We refer to the survey \cite{broadie2004anniversary} for details about extensions of the BSM model, other classical option pricing models, and numerical methods such as the Monte Carlo method.

In a complete market one can \emph{hedge} a given derivative contract by buying and selling the underlying asset in the right way to eliminate risk. In the Black-Scholes analysis \emph{delta} hedging is used, in which we hedge the risk of a call option by shorting $\Delta_t$ units of the underlying asset with $\Delta_t:=\frac{\partial V}{\partial S}(S_t,t)$ (the sensitivity of the option price with respect to the asset price). It is also possible to use financial derivatives to \emph{hedge} against the volatility of given positions in the underlying assets. However, in practice, we can only rebalance portfolios at discrete time points and frequent transactions may incur high costs. Therefore an optimal hedging strategy depends on the tradeoff between the hedging error and the transaction costs. It is worth mentioning that this is in a similar spirit to the mean-variance portfolio optimization framework introduced in Section \ref{sec:portfolio_optimization}. Much effort has been made to include the transaction costs using classical approaches such as dynamic programming, see, for example \cite{leland1985option,figlewski1989options,henrotte1993transaction}. We also refer to \cite{giurca2021} for the details about delta hedging under different model assumptions. 

However, the above discussion addresses option pricing and hedging in a model-based way since there are strong assumptions made on the asset price dynamics and the form of transaction costs.  In practice some assumptions of the BSM model are not realistic since: 1) transaction costs due to commissions, market impact, and non-zero bid-ask spread exist in the real market; 2) the volatility is not constant; 3) short term returns typically have a heavy tailed distribution \cite{cont2001empirical,chakraborti2011econophysics}.   Thus the resulting prices and hedges may suffer from model mis-specification when the real asset dynamics is not exactly as assumed and the transaction costs are difficult to model. Thus we will focus on a model-free RL approach that can address some of these issues.

\paragraph{RL Approach.} RL methods including (deep) $Q$-learning \cite{halperin2020bs,du2020option,cao2021hedging,li2009,halperin2019qlbs}, PPO \cite{du2020option}, and DDPG \cite{cao2021hedging} have been applied to find hedging strategies and price financial derivatives. The state variables often include asset price, current positions, option strikes, and time remaining to expiry. The control variable is often set to be the change in holdings. Examples of reward functions are (risk-adjusted) expected wealth/return \cite{halperin2020bs,halperin2019qlbs,du2020option,kolm2019} (as in the mean-variance portfolio optimization), option payoff \cite{li2009}, and (risk-adjusted) hedging cost \cite{cao2021hedging}. The benchmarks for pricing models are typically the BSM model \cite{halperin2020bs,halperin2019qlbs} and binomial option pricing model \cite{dubrov2015} (introduced in \cite{cox1979option}). The learned hedging strategies are typically compared to a discrete approximation to the delta hedging strategy for the BSM model  \cite{cao2021hedging,du2020option,buehler2019hedging,cannelli2020hedging}, or the hedging strategy for the Heston model \cite{buehler2019hedging}.  In contrast to the BSM model where the volatility is assumed to be constant, the Heston model assumes that the volatility of the underlying asset follows a particular stochastic process which leads to a semi-analytical solution. The performance measures for the hedging strategy used in RL papers include (expected) hedging cost/error/loss \cite{cao2021hedging,kolm2019,buehler2019hedging,cannelli2020hedging}, PnL \cite{du2020option,kolm2019,cannelli2020hedging}, and average payoff \cite{li2009}. Some practical issues have been taken into account in RL models, including transaction costs \cite{buehler2019hedging,du2020option,cao2021hedging,kolm2019} and position constraints such as round lotting \cite{du2020option} (where a round lot is a standard number of securities to be traded, such as 100 shares) and limits on the trading size (for example buy or sell up to 100 shares) \cite{kolm2019}.

For European options, \cite{halperin2020bs} developed a discrete-time option pricing model called the QLBS ($Q$-Learner in Black-Scholes) model based on Fitted $Q$-learning (see Section \ref{subsec:deep_value_based}). This model learns both the option price and the hedging strategy in a similar spirit to the mean-variance portfolio optimization framework.  \cite{halperin2019qlbs} extended the QLBS model in \cite{halperin2020bs} by using Fitted $Q$-learning. They also investigated the model in a different setting where the agent infers the risk-aversion parameter in the reward function using observed states and actions. However, \cite{halperin2020bs} and \cite{halperin2019qlbs} did not consider transaction costs in their analysis. By contrast, \cite{buehler2019hedging} used deep neural networks to approximate an optimal hedging strategy under market frictions, including transaction costs, and \emph{convex risk measures} \cite{kloppel2007dynamic} such as conditional value at risk. They showed that their method can accurately recover the optimal hedging strategy in the Heston model \cite{Heston1993} without transaction costs and it can be used to numerically study the impact of proportional transaction costs on option prices. The learned hedging strategy is also shown to outperform delta hedging in the (daily recalibrated) BSM model in a simple setting, this is hedging an at-the-money European call option on the S\&P 500 index. The method in \cite{buehler2019hedging} was extended in \cite{carbonneau2021} to price and hedge a very large class of derivatives including
vanilla options and exotic options in more complex environments (e.g. stochastic volatility models and jump processes). \cite{kolm2019} found the optimal hedging strategy by optimizing a simplified version of the mean-variance objective function \eqref{eqn:MV_obj}, subject to discrete trading, round lotting, and nonlinear transaction costs. They showed in their simulations that the learned hedging strategy achieves a much lower cost, with no significant difference in volatility of total PnL, compared to the delta hedging strategy. \cite{du2020option} extended the framework in \cite{kolm2019} and tested the performance of DQN and PPO for European options with different strikes. Their simulation results showed that these models are superior to delta hedging in general, and out of all models, PPO achieves the best performance in terms of PnL, training time, and the amount of data needed for training.  \cite{cannelli2020hedging} formulated the optimal hedging problem as a Risk-averse Contextual $k$-Armed Bandit (R-CMAB) model (see the discussion of the contextual bandit problem in Section \ref{sec:MDP2learning}) and proposed a deep CMAB algorithm involving Thompson Sampling \cite{thompson1933likelihood}. They showed that their algorithm outperforms DQN in terms of sample efficiency and hedging error when compared to delta hedging (in the setting of the BSM model). Their learned hedging strategy was also shown to converge to delta hedging in the absence of risk adjustment, discretization error and transaction costs. \cite{cao2021hedging} considered $Q$-learning and DDPG for the problem of hedging a short position in a call option when there are transaction costs. The objective function is set to be a weighted sum of the expected hedging cost and the standard deviation of the hedging cost. They showed that their approach achieves a markedly lower expected hedging cost but with a slightly higher standard deviation of the hedging cost when compared to delta hedging. In their simulations, the stock price is assumed to follow either geometric Brownian motion or a stochastic volatility model.

For American options, the key challenge is to find the optimal exercise strategy which determines when to exercise the option as this determines the price. \cite{li2009} used the Least-Squares Policy Iteration (LSPI) algorithm \cite{lagoudakis2003least} and the Fitted $Q$-learning algorithm to learn the exercise policy for American options. In their experiments for American put options using both real and synthetic data, the two algorithms gain larger average payoffs than the benchmark Longstaff-Schwartz method \cite{longstaff2001}, which is the standard Least-Squares Monte Carlo algorithm. \cite{li2009} also analyzed their approach from a theoretical perspective and derived a high probability, finite-time bound on their method. \cite{dubrov2015} then extended the work in  \cite{li2009} by combining random forests, a popular machine learning technique, with Monte Carlo simulation for pricing of both American options and \emph{convertible bonds}, which are corporate bonds that can be converted to the stock of the issuing
company by the bond holder. They showed that the proposed algorithm provides more accurate prices than several other methods including LSPI, Fitted $Q$-learning, and the Longstaff-Schwartz method.


\subsection{Market Making}
A market maker in a financial instrument is an individual trader or an institution that provides liquidity to the market 
by placing buy and sell limit orders in the LOB for that instrument while earning the bid-ask spread.

The objective in market making is different from problems in optimal execution (to target a position) or portfolio optimization (for long-term investing). Instead of profiting from identifying the  correct price movement direction, the objective of a market maker is to profit from earning the bid-ask spread without accumulating undesirably large  positions (known as inventory) \cite{gueant2012optimal}. A market maker faces three major sources of risk \cite{guilbaud2013optimal}. The inventory risk \cite{avellaneda2008high} refers to the risk of accumulating an undesirable large net inventory, which significantly increases volatility due to market movements. The execution risk \cite{kuhn2010optimal} is the risk that limit orders may not get filled over a desired horizon. Finally, the adverse selection risk refers to the situation where there is a directional price movement that sweeps through the limit orders submitted by the market marker such that the price does not revert back by the end of the trading horizon. This may lead to a huge loss as the market maker in general needs to clear their inventory at the end of the horizon (typically the end of the day to avoid overnight inventory).

\paragraph{Stochastic Control Approach.} Traditionally the theoretical study of market making strategies follows a stochastic control approach, where the LOB dynamics are modeled directly by some stochastic process and an optimal market making strategy that maximizes the market maker's expected utility can be obtained by solving the Hamilton--Jacobi--Bellman equation. See \cite{avellaneda2008high}, \cite{gueant2013dealing} and \cite{obizhaeva2013optimal}, and 
\cite[Chapter 10]{cartea2015algorithmic} for examples.

We follow the framework in \cite{avellaneda2008high} as an example to demonstrate the control formulation in continuous time and discuss its advantages and disadvantages.

Consider a high-frequency market maker trading on a single stock over a finite horizon $T$. Suppose the mid-price of this stock follows an arithmetic Brownian motion in that
\begin{eqnarray}\label{eq:MM_dynamics}
d S_t = \sigma d W_t,
\end{eqnarray}
where $\sigma$ is a constant and $W = \{W_t\}_{0\leq t \leq T}$ is a standard Brownian motion defined on a filtered probability space $(\Omega,\mathcal{F},\{\mathcal{F}\}_{0\leq t \leq T},\mathbb{P})$.

The market maker will continuously propose bid and ask prices,
denoted by $S^b_t$ and $S^a_t$ respectively, and hence will buy and sell shares according to the rate
of arrival of market orders at the quoted prices. Her inventory, which is the number of shares she holds, is therefore given by
\begin{eqnarray}
q_t = N_t^b - N_t^a,
\end{eqnarray}
where $N^b$ and $N^a$ are the point processes (independent of $W$) giving the number of shares the
market maker respectively bought and sold. Arrival rates obviously depend on the prices $S^b_t$ and $S^a_t$ quoted by the market maker and we assume that the intensities $\lambda^b$ and $\lambda^a$ associated respectively to $N^b$ and $N^a$ depend on the difference between the quoted prices and the reference price (i.e. $\delta^b_t = S_t - S^b_t$ and $\delta^a_t = S_t - S^a_t$)
 and are of the following form
\begin{eqnarray}
\lambda^b(\delta^b)=A\exp(-k\delta^b) \mbox{ and } \lambda^a(\delta^a)=A\exp(-k\delta^a), 
\end{eqnarray}
where $A$ and $k$ are positive constants that characterize the liquidity of the stock. Consequently the cash process of the market maker follows
\begin{eqnarray}
d X_t = (S_t+\delta_t^a) d N_t^a -  (S_t-\delta_t^b) d N_t^b.
\end{eqnarray}
 Finally, the market maker optimizes a constant absolute risk aversion (CARA) utility function:
\begin{eqnarray}
V(s,x,q,t) = \sup_{\{\delta_u^a,\delta_{u}^b\}_{t\leq u \leq T}\in \mathcal{U}} \mathbb{E} \bigg[ -\exp (-\gamma(X_T+q_T S_T))\bigg| X_t = x, S_t = s, \mbox{ and } q_t = q\bigg],
\end{eqnarray}
where $\mathcal{U}$ is the set of predictable processes bounded from below, $\gamma$ is the absolute risk aversion coefficient characterizing the market maker, $X_T$ is the amount of cash at time $T$ and
$q_T S_T$ is the evaluation of the (signed) remaining quantity of shares in the inventory at time $T$. 

By applying the dynamic programming principle, the value function $V$ solves the following Hamilton--Jacobi--Bellman equation:
\begin{eqnarray}\label{avalaneda-hjb}
\begin{cases}
&\partial_t V +\frac{1}{2} \partial_{ss} V + \max_{\delta^b} \lambda^b (\delta^b) [V(s,x-s+\delta^b,q+1,t) - u(s,x,q,t)]\\
&\qquad\qquad\qquad + \max_{\delta^a} \lambda^a (\delta^a) [V(s,x+s+\delta^a,q-1,t) - u(s,x,q,t)] = 0,\\
&V(s,x,q,T) = -\exp(-\gamma(x+qs)).
\end{cases}
\end{eqnarray}

While \eqref{avalaneda-hjb}, derived in \cite{avellaneda2008high},
admits a (semi) closed-form solution which leads to nice insights about the problem, it all builds on the full analytical specification of the market dynamics. In addition, there are very few utility functions (e.g. exponential (CARA), power (CRRA), and  quadratic) known in the literature that could possibly lead to closed-form solutions. The same issues arise in other work along this line (\cite{gueant2013dealing} and \cite{obizhaeva2013optimal}, and \cite[Chapter 10]{cartea2015algorithmic}) where strong model assumptions are made about the prices or about the LOB or both. This requirement of full analytical specification means these papers are quite removed from realistic market making, as financial markets do not conform to any simple parametric model specification with fixed parameters. 

\paragraph{RL Approach.} For market making problems with an RL approach, 
both value-based methods (such as the $Q$-learning algorithm  \cite{abernethy2013adaptive,spooner2018market} and SARSA \cite{spooner2018market}) and policy-based methods (such as deep policy gradient method \cite{zhao2021high}) have been used. The state variables are often composed of bid and ask prices, current holdings of assets, order-flow imbalance, volatility, and some sophisticated market indices. The control variables are typically set to be the spread to post a pair of limit buy and limit sell orders. Examples of reward include PnL with inventory cost  \cite{abernethy2013adaptive,spooner2018market,spooner2020robust,zhao2021high} or Implementation Shortfall with inventory cost \cite{gavsperov2021market}.

The first RL method for market making
was explored by \cite{chan2001electronic} and the authors applied three RL algorithms and tested them in simulation environments: a Monte Carlo method, a SARSA method and an actor-critic method. 
For all three methods, the state variables include inventory of the market-maker, order imbalance on the market, and market quality measures.  The actions are the changes in the bid/ask price to post the limit orders and the sizes of the limit buy/sell orders. The reward function is set as a linear combination of profit (to maximize), inventory risk (to minimize), and market qualities (to maximize). The authors found that SARSA and Monte Carlo methods work well in a simple simulation environment. The actor-critic method is more plausible in complex environments and generates stochastic policies that correctly adjust bid/ask prices with respect to order imbalance and effectively control the trade-off between the profit and the quoted spread (defined as the price difference to post limit buy orders and limit sell orders). Furthermore, the stochastic policies are shown to outperform deterministic policies in achieving a lower variance of the resulting spread. Later on, \cite{abernethy2013adaptive} designed a group of ``spread-based'' market making strategies parametrized by a minimum quoted spread. The strategies bet on the mean-reverting behavior of the mid-price and utilize the opportunities when the mid-price deviates from the price during the previous period. Then an online algorithm is used to pick in each period a minimum quoted spread. The states of the market maker are the current inventory and price data. The actions are the quantities and at what prices to offer in the market. It is assumed that the market maker interacts with a continuous double auction via an order book. The market maker can place both market and limit orders and is able to make and cancel orders after every price fluctuation. The authors provided structural properties of these strategies which allows them to obtain low regret relative to the best such strategy in hindsight which maximizes the realized rewards. 
\cite{spooner2018market} generalized the results in \cite{chan2001electronic} and adopted several reinforcement learning algorithms (including $Q$-learning and SARSA) to improve the decisions of the market maker. In their framework, the action space contains  ten actions. The first nine actions correspond to a pair of orders with
a particular spread and bias in their prices. The final action  allows
the agent to clear their inventory using a market order. The states can be divided into two groups: agent states and market states. Agent states include inventory level and active quoting distances and market states include market (bid-ask) spread, mid-price move, book/queue imbalance, signed volume, volatility, and relative strength index. The reward function is designed as the sum of a symmetrically dampened PnL and an inventory cost.  The idea of the symmetric damping is to
disincentivize the agent from trend chasing and direct the agent towards spread capturing. A simulator of a financial market is constructed via direct reconstruction of the LOB from historical data. Although, since the market is reconstructed from historical data, simulated orders placed by an agent cannot impact the market. The authors compared their algorithm to a modified version of \cite{abernethy2013adaptive} and showed a significant empirical improvement from \cite{abernethy2013adaptive}. To address the adverse selection risk that market makers are often faced with in a high-frequency environment, \cite{zhao2021high}  proposes
a high-frequency feature Book Exhaustion Rate (BER) and shows
theoretically and empirically that the BER can serve as a direct measurement of the adverse selection risk from an equilibrium point
of view. The authors train a market making algorithm via a deep policy-based RL algorithm using three years of LOB data for the Chicago Mercantile Exchange (CME) S\&P 500 and 10-year Treasury note futures. The numerical performance demonstrates that utilizing the BER allows the algorithm to avoid large losses due to adverse selection and achieve stable performance.

Some recent papers have focused on improving the robustness of  market markers' strategies with respect to adversarial and
volatile market conditions. \cite{gavsperov2021market} used perturbations by an opposing agent - the adversary - to render the market maker  more robust to
model uncertainty and consequently improve generalization. Meanwhile, \cite{gavsperov2021market} incorporated additional predictive signals in the states to improve the learning outcomes of market makers. In particular, the states include the
agent's inventory,  price range and trend predictions. In a similar way to the setting in \cite{abernethy2013adaptive} and \cite{spooner2018market}, actions are represented by a certain choice of bid/ask orders
relative to the current best bid/ask. The reward function both incentivizes spread-capturing (making round-trips) and discourages holding inventory. The first part of the reward is inspired by utility functions containing a running inventory-based penalty with an absolute value inventory penalty term which has a convenient Value at Risk (VAR) interpretation. The second part is a symmetrically dampened PnL which follows \cite{spooner2018market}. Experimental results on historical data demonstrate the superior reward-to-risk performance of the proposed framework over several standard market making benchmarks. More specifically, in these experiments, the resulting reinforcement learning agent achieves between 20-30\% higher
terminal wealth than the benchmarks while being exposed to only around 60\% of their inventory risks.   Also,
 \cite{spooner2020robust} applied
 Adversarial Reinforcement Learning
(ARL)  to a zero-sum game version of the control formulation  \eqref{eq:MM_dynamics}-\eqref{avalaneda-hjb}. The adversary acts as a proxy for other market participants that would like to profit at the
market maker's expense. 

In addition to designing learning algorithms for the market maker in an unknown financial environment, RL algorithms can also be used to solve high-dimensional control problems for the market maker or to solve the control problem with the presence of a time-dependent rebate in the full information setting. In particular, \cite{gueant2019deep} focused on a setting where a market maker needs to decide the optimal bid and ask quotes for a given universe of bonds in an OTC market. This problem is high-dimensional and  other classical numerical methods including finite differences are inapplicable.  The authors proposed  a
model-based Actor-Critic-like algorithm involving a deep neural network to numerically solve the problem. Similar ideas have been applied to market making problems in dark pools \cite{baldacci2019market} (see the discussion on dark pools in Section \ref{sec:SOR}). With the presence of a time-dependent rebate, there is no closed-form solution for the associated stochastic control problem of the market maker. Instead, \cite{zhang2020reinforcement} proposed a Hamiltonian-guided value function approximation algorithm to solve for the numerical solutions under this scenario.

Multi-agent RL algorithms are also used to improve the strategy for market making with a particular focus on the impact of competition from other market makers or the interaction with other types of market participant. See \cite{ganesh2019reinforcement} and \cite{patel2018optimizing}.

\subsection{Robo-advising}
\label{sec:robo-advising}
Robo-advisors, or automated investment managers, are a class of financial advisers that provide online financial advice or investment management  with  minimal human intervention. They provide digital financial advice based on mathematical rules or algorithms which can easily take into account different sources of data such as news, social media information, sentiment data and earnings reports. Robo-advisors  have gained widespread popularity  and  emerged prominently as an alternative to traditional human advisers in recent years.  The first robo-advisors were launched after the 2008 financial crisis when financial services institutions were facing the ensuing loss of trust from their clients. Examples of pioneering robo-advising firms include Betterment and Wealthfront. As of 2020, the value of assets under robo-management is highest in the United States and exceeded \$650 billion \cite{capponi2021personalized}.

The robo-advisor does not know the client's risk preference in advance but learns it while interacting with the client. The robo-advisor then improves its investment
decisions based on its current estimate of the client's risk preference. There are several challenges in the application of robo-advising. Firstly, the client's risk preference may change over-time and may depend on the market returns and economic conditions. Therefore the robo-advisor needs to determine a frequency of interaction with the client that ensures a high level of consistency in the risk preference when adjusting portfolio allocations. Secondly, the robo-advisor usually faces a dilemma when it comes to either catering to the client's wishes, that is, investing according to the client's risk preference, or going against the client's wishes in order to seek better investment performance. Finally, there is also a subtle trade-off between the rate of information acquisition from the client and the accuracy of the acquired information.
On the one hand, if the interaction does not occur at all times, the robo-advisor may not always have access to up-to-date information about the client's profile. On the other hand, information communicated to the robo-advisor may not be representative of the client's true risk aversion as the client is subject to behavioral biases.

\paragraph{Stochastic Control Approach.} To address the above mentioned challenges, \cite{capponi2021personalized} proposed a stochastic control framework with four components: (i) a regime switching model of market returns, (ii) a mechanism of interaction between the client and the robo-advisor, (iii) a dynamic model (i.e., risk aversion process) for the client's risk preferences, and (vi) an optimal investment criterion. In this framework, the robo-advisor interacts repeatedly with the client and learns about changes in her risk profile whose evolution is specified by (iii). The robo-advisor adopts a multi-period mean-variance investment criterion with a finite investment horizon based on the estimate of the client's risk aversion level. The authors showed the stock market allocation resulting from the dynamic mean-variance optimization consists of two components where the first component is akin to the standard single period Markowitz strategy whereas the second component is intertemporal hedging demand, which depends on the relationship between the current market return and future portfolio returns. 

Note that although \cite{capponi2021personalized} focused on the stochastic control approach to tackle the robo-advising problem, the framework is general enough that some components (for example the mean-variance optimization step) may be replaced by an RL algorithm and the dependence on model specification can be potentially relaxed.

\paragraph{RL Approach.} There are only a few references on robo-advising with an RL approach since this is still a relatively new topic. We  review each paper with details. The first RL algorithm for a robo-advisor was proposed by
\cite{alsabah2021robo} where the authors designed an exploration-exploitation algorithm to learn the investor's risk appetite over time by observing her portfolio choices in different market environments. The set of various market environments of interest is formulated as the state space $\mathcal{S}$. In each period, the robo-advisor places an investor's capital into one of several pre-constructed portfolios which can be viewed as the action space $\mathcal{A}$. Each portfolio decision reflects the robo-advisor's belief concerning the investor's true risk preference from a discrete set of possible risk aversion parameters $
\Theta = \{\theta_i\}_{1\leq i \leq |\Theta|}$. The investor interacts with the robo-advisor by portfolio selection choices, and such interactions are used to update
the robo-advisor's estimate of the investor's risk profile.  
The authors proved that, with high probability, the proposed exploration-exploitation algorithm performs near optimally with the number of time steps depending polynomially on various model parameters.

\cite{wang2021robo} proposed an investment robo-advising framework consisting of two agents. The first
agent, an inverse portfolio optimization agent, infers an investor's risk preference and expected return directly from historical allocation data using online inverse optimization. The
second agent, a deep RL agent, aggregates the inferred sequence of expected returns to formulate
a new multi-period mean-variance portfolio optimization problem that can be solved using a deep RL approach based on the DDPG method. The proposed investment pipeline was applied to real market data from
April 1, 2016 to February 1, 2021 and was shown to consistently outperform the S\&P 500 benchmark portfolio that represents the aggregate market optimal allocation. 

As mentioned earlier in this subsection, learning the client's risk preference is challenging as the preference may depend on multiple factors and may change over time. \cite{yu2020learning} was dedicated to learning the risk preferences from investment portfolios using an inverse optimization technique. In particular, the proposed inverse optimization approach can be used to measure time varying risk preferences directly from market signals and portfolios. This approach is developed based on two methodologies: convex optimization based modern portfolio theory and learning the decision-making scheme through inverse optimization. 

\subsection{Smart Order Routing}\label{sec:SOR} In order to execute a trade of a given asset, market participants  may have the opportunity to split the trade and submit orders to different venues, including both lit pools and dark pools, where this asset is traded. This could potentially improve the overall execution price and quantity. Both the decision and hence the outcome are influenced by the characteristics of different venues as well as the structure of transaction fees and rebates across different venues.

\paragraph{Dark Pools vs. Lit Pools.} Dark pools are private exchanges for trading securities that are not accessible by the investing public. Also known as ``dark pools of liquidity'', the name of these exchanges is a reference to their complete lack of transparency. Dark pools were created in order to facilitate block trading by institutional investors who did not wish to impact the markets with their large orders and obtain adverse prices for their trades. According to recent Securities and Exchange Commission (SEC) data, there were 59 registered Alternative Trading Systems (a.k.a. the ``Dark Pools'') with the SEC as of May 2021 of which there are three types: (1) Broker-Dealer-Owned Dark Pools, (2) Agency Broker or Exchange-Owned Dark Pools, and (3) Electronic Market Makers Dark Pools. Lit pools are effectively the opposite of dark pools. Unlike dark pools, where prices at which participants are willing to trade are not revealed, lit pools do display bid offers and ask offers in different stocks. Primary exchanges operate in such a way that available liquidity is displayed at all times and form the bulk of the lit pools available to traders.

For smart order routing (SOR) problems, the most important characteristics of different dark pools are the chances of being  matched with a counterparty and the price (dis)advantages whereas the relevant characteristics of lit pools include the order flows, queue sizes, and cancellation rates.

There are only a few references on using a data-driven approach to tackle SOR problems  for dark pool allocations and for allocations across lit pools. We will review each of them with details.

\paragraph{Allocation Across Lit Pools.} The SOR problem across multiple lit pools (or primary exchanges) was first studied in \cite{cont2017optimal} where the authors formulated the SOR problem as a convex optimization problem.

Consider a trader who needs to buy $S$ shares of a stock within a short time interval $[0, T]$. At time $0$, the trader may submit $K$ limit orders with $L_k \ge 0$ shares to exchanges $k = 1,\cdots,K$ (joining the queue of the best bid price level) and also market orders for $M \ge 0$ shares.
At time $T$ if the total executed quantity is less than $S$ the trader also submits a market order to execute the remaining amount. The trader's order placement decision is thus summarized by a
vector $X=(M,L_1,\cdots,L_K) \in \mathbb{R}^{K+1}_+$ of order sizes.
It is assumed for simplicity that a market order of any size up to $S$ can be filled immediately at
any single exchange. Thus a trader chooses the cheapest venue for his market orders. 

Limit orders with quantities $(L_1,\cdots, L_K)$ join queues of $(Q_1,\cdots, Q_K)$ orders in $K$ limit order
books, where $Q_k \ge 0$.  Then the filled amounts at the end of the horizon $T$ can be written as a function of their initial queue position and future order flow:
\begin{eqnarray}\label{eq:firfull_amnt}
\min(\max(\xi_k - Q_k, 0), L_k) = (\xi_k-Q_k)_+ - (\xi_k - Q_k - L_k)_+
\end{eqnarray}
where the notation $(x)_+ = \max(0,x)$. Here $\xi_k:=D_k+C_k$ is the total outflow from the front of the $k$-th queue which consists of
order cancellations $C_k$ that occurred before time $T$ from queue positions in front of an order and market orders $D_k$ that reach the $k$-th exchange before $T$.  Note that the fulfilled amounts are random because they depend on queue outflows $\xi=(\xi_1,\cdots,\xi_K)$ during $[0, T]$, which are modeled as random variables with a distribution $F$.

Using the mid-quote price as a benchmark, the execution cost relative to the mid-quote for an order allocation $X = (M, L_1,.. ., L_K)$ is defined as:
\begin{eqnarray}
V_{\rm execution}(X, \xi) := (h + f)M -\sum_{k=1}^K (h + r_k)((\xi_k - Q_k)_+ - (\xi_k - Q_k - L_k)_+),
\end{eqnarray}
where $h$ is one-half of the bid-ask spread at time $0$, $f$ is a fee for market orders and $r_k$ are effective rebates for providing liquidity by submitting limit orders on exchanges $k = 1,\cdots,K$.

Penalties for violations of the target quantity in both directions are included:
\begin{eqnarray}
V_{\rm penalty}(X, \xi) := \lambda_u (S-A(X, \xi))_{+}+\lambda_o (A(X, \xi)-S)_{+},
\end{eqnarray}
where $\lambda_o \ge 0$ and $\lambda_u \ge 0$ are, respectively, the penalty for overshooting and undershooting, and $A(X,\xi) = M+ \sum_{k=1}^K (h + r_k)((\xi_k - Q_k)_+ - (\xi_k - Q_k - L_k)_+)$ is the total number of shares bought by the trader during $[0, T)$.

The impact cost is paid on
all orders placed at times 0 and $T$, irrespective of whether they are filled, leading to the following total impact:
\begin{eqnarray}
V_{\rm impact}(X,\xi) = \theta (M+L_k +(S-A(X,\xi)_+))
\end{eqnarray}
where $\theta > 0$ is the impact coefficient.

Finally, the cost function is defined as the summation of all three pieces $V(X,\xi): = V_{\rm execution}(X, \xi)+ V_{\rm penalty}(X, \xi) + V_{\rm impact}(X,\xi)$. \cite[Proposition 4]{cont2017optimal}  provides  optimality conditions for an order allocation $X^* = (M^*,L_1,\cdots,L_K)$. In particular,  (semi)-explicit model conditions are given for when $L_k^*>0$ ($M^*>0$), i.e., when submitting limit orders to venue $k$ (when submitting market orders)  is optimal.

In a different approach to the single-period model introduced above, \cite{baldacci2020adaptive} formulated the SOR problem as an order allocation problem across multiple lit pools over multiple trading periods. Each venue is characterized by a bid-ask spread process and  an imbalance process. The dependencies between the imbalance and spread at the venues are considered through a covariance matrix. A Bayesian learning framework for learning and updating the model parameters is proposed to take into account possibly changing market conditions. Extensions to include short/long trading signals, market impact or hidden liquidity are also discussed.

\paragraph{Allocation Across Dark Pools.} As discussed at the beginning of Section \ref{sec:SOR}, dark pools are a type of stock exchange that is designed to facilitate large transactions. A
key aspect of dark pools is the {\it censored feedback} that
the trader receives. At every iteration the trader has a
certain number $V_t$ of shares to allocate amongst $K$
different dark pools with $v_t^i$ the volume allocated to dark pool $i$.  The dark pool $i$ trades as many
of the allocated shares $v_t^i$ as it can with the available
liquidity $s_t^i$. The trader only finds out how many of these
allocated shares were successfully traded at each dark
pool (i.e., $\min(s_t^i,v_t^i)$), but not how many would have been traded if
more were allocated (i.e., $s_t^i$).  

Based on this property of censored feedback, \cite{ganchev2010censored} formulated the allocation problem across dark pools as an online learning problem under the assumption that $s_t^i$ is an i.i.d. sample from some unknown distribution $P_i$ and the total allocation quantity $V_t$ is an i.i.d. sample from an unknown distribution $Q$ with $V_t$ upper bounded by a constant $V>0$ almost surely. At each iteration $t$, the learner allocates the orders greedily according to the  estimate $\widehat{P}^{(t-1)}_i$ of the distribution $P_i$ for all dark pools $i=1,2,\cdots, K$ derived from previous iteration $t-1$. Then the learner can update the estimation $\widehat{P}^{(t)}_i$ of the distribution $P_i$ with a modified version of the Kaplan-Meier estimate (a non-parametric statistic used to estimate the cumulative probability) with the new censored observation $\min(s_t^i,v_t^i)$ from iteration $t$. The authors then proved that for any $\varepsilon > 0$ and
$ \delta> 0$, with probability $1-\delta$ (over the randomness
of draws from Q and $\{P_i\}_{i=1}^K$), after running for a time
polynomial in $K, V , 1/\varepsilon$, and $\ln(1/\delta)$, the algorithm
makes an $\varepsilon$-optimal allocation on each subsequent time
step with probability at least $1-\varepsilon$.

The  setup of \cite{ganchev2010censored} was generalized in \cite{agarwal2010optimal} where the authors assumed that the sequences of volumes $V_t$ and available liquidities $\{s_t^i\}_{i=1}^K$ are chosen by an adversary who knows the previous allocations of their algorithm.
An exponentiated gradient style algorithm was proposed and shown to enjoy an optimal regret guarantee $\mathcal{O}(V\sqrt{T \ln K})$ against the best allocation strategy in hindsight.

\section{Further Developments for Mathematical Finance and Reinforcement Learning}

The general RL algorithms developed in the machine learning literature are good starting points for use in financial applications. A possible drawback is that such general RL algorithms tend to overfit, using more information than is actually required for a particular application. On the other hand, the stochastic control approach to many financial decision-making problems may suffer from the risk of model mis-specification. However, it may capture the essential features of a given financial application from a modeling perspective, in terms of the dynamics and the reward function. 

One promising direction for RL in finance is to develop an even closer integration of  the modeling techniques (the domain knowledge) from the stochastic control literature and key components of a given financial application (for example the adverse selection risk for market-making problems and the execution risk for optimal liquidation problems) with the learning power of the RL algorithms. This line of developing a more integrated framework is interesting from both theoretical and applications perspectives. From the application point of view, a modified RL algorithm, with designs tailored to one particular financial application, could lead to better empirical performance. This could be verified by comparison with existing algorithms on the available datasets. In addition, financial applications motivate potential new frameworks and playgrounds for RL algorithms. Carrying out the convergence and sample complexity analysis for these modified algorithms would also be a meaningful direction in which to proceed. Many of the papers referenced in this review provide great initial steps in this direction. We list the following future directions that  the reader may find interesting. 
 
\paragraph{Risk-aware or Risk-sensitive RL.} Risk arises from the uncertainties associated with future events, and is inevitable since the consequences of actions are uncertain at the time when a decision is made. Many decision-making problems in finance lead to trading strategies and it is important to account for the risk of the proposed strategies (which could be measured for instance by the maximum draw-down, the variance or the 5\% percentile of the PnL distribution) and/or the risk from the market environment such as the adverse selection risk.

Hence it would be interesting to include risk measures in the design of RL algorithms for financial applications.  The challenge of risk-sensitive RL lies both in the non-linearity of the objective function with respect to the reward and in
designing a risk-aware exploration mechanism. 

RL with risk-sensitive utility functions has been studied in several papers without regard to specific financial applications.  The work of \cite{mihatsch2002risk} proposes TD(0) and $Q$-learning-style algorithms that transform temporal differences
instead of cumulative rewards, and proves their convergence. Risk-sensitive RL with a general
family of utility functions is studied in \cite{shen2014risk}, which also proposes a $Q$-learning algorithm with convergence guarantees. The work of \cite{eriksson2019epistemic} studies a risk-sensitive policy gradient algorithm, though
with no theoretical guarantees. \cite{fei2020risk}  considers the problem of risk-sensitive RL with exponential utility and proposes
two efficient model-free algorithms, Risk-sensitive Value Iteration (RSVI) and Risk-sensitive $Q$-learning (RSQ), with a near-optimal sample complexity guarantee. 
\cite{vadori2020risk} developed a martingale approach to  learn policies that are sensitive to the uncertainty of the rewards and are meaningful under some market scenarios.
Another line of work focuses on constrained RL problems with different risk criteria \cite{achiam2017constrained,chow2017risk,chow2015risk,ding2021provably,tamar2015policy,zheng2020constrained}.  Very recently, \cite{jaimungal2021robust} proposed a robust risk-aware reinforcement learning framework via robust optimization and with a rank dependent expected utility function. Financial applications such as statistical arbitrage and portfolio optimization are discussed  with detailed numerical examples. \cite{coache2021reinforcement}  develops a framework combining policy-gradient-based RL method and dynamic convex risk measures for solving time-consistent risk-sensitive stochastic optimization problems. However, there is no sample complexity or asymptotic convergence studied for the proposed algorithms in \cite{jaimungal2021robust,coache2021reinforcement}. 

\paragraph{Offline Learning and Online Exploration.}
Online learning requires updating of algorithm parameters in real-time and this is impractical for many financial decision-making problems, especially in the high-frequency regime. The most plausible setting is to collect data with a pre-specified exploration scheme during trading hours and update the algorithm with the new collected data after the close of trading. This is closely 
related to the translation of online learning to offline regression \cite{simchi2020bypassing} and RL with batch data \cite{chen2019information,gao2019batched,garcelon2020conservative,ren2020dynamic}. However, these  developments focus on general methodologies without being specifically tailored to financial applications.

\paragraph{Learning with a Limited Exploration Budget.} Exploration can help agents to find new policies to improve their future cumulative rewards. However, too much exploration can be both time consuming and computation consuming, and in particular, it may be very costly for some financial applications. Additionally, exploring black-box trading strategies may need a lot of justification within a financial institution and hence investors tend to limit the effort put into exploration and try to improve performance as much as possible within a given budget for exploration.  This idea is similar in spirit to conservative RL where agents explore new strategies to maximize revenue whilst simultaneously maintaining their revenue above a fixed baseline, uniformly over time \cite{wu2016conservative}. This is also related to the problem of information acquisition with a cost which has been studied for economic commodities \cite{pomatto2018cost} and operations management \cite{ke2016search}. It may also be interesting to investigate such costs for decision-making problems in financial markets.

\paragraph{Learning with Multiple Objectives.} In finance, a common problem is to choose a portfolio when there are two conflicting objectives - the desire to have the expected value of portfolio returns be as high as possible, and the desire to have risk, often measured by the standard deviation of portfolio returns, be as low as possible. This problem is often represented by a graph in which the efficient frontier shows the best combinations of risk and expected return that are available, and in which indifference curves show the investor's preferences for various risk-expected return combinations. Decision makers sometimes combine both criteria into a single objective function consisting of the difference of the expected reward and a scalar multiple of the risk. However, it may well not be in the best interest of a decision maker to combine relevant criteria in a linear format for certain applications. For example, market makers on the OTC market tend to view criteria such as turn around time, balance sheet constraints, inventory cost, profit and loss as separate objective functions. The study of multi-objective RL is still at a preliminary stage and relevant references include \cite{zhou2020provable} and \cite{yang2019generalized}.


\paragraph{Learning to Allocate Across Lit Pools and Dark Pools.} Online optimization methods explored in \cite{agarwal2010optimal} and \cite{ganchev2010censored} for dark pool allocations can be viewed as a single-period RL algorithm and the Bayesian framework developed in \cite{baldacci2020adaptive} for allocations across lit pools may be classified as a model-based RL approach. However, there is currently no existing work on applying multi-period and model-free RL methods to learn how to route orders across both dark pools and lit pools. We think this might be an interesting direction to explore as  agents sometimes have access to both lit pools and dark pools and these two contrasting pools have quite different information structures and matching mechanisms.

\paragraph{Robo-advising in a Model-free Setting.}
As introduced in Section \ref{sec:robo-advising}, \cite{alsabah2021robo} considered learning  within a set of $m$ pre-specified investment portfolios, and \cite{wang2021robo}  and \cite{yu2020learning} developed  learning algorithms and procedures to infer risk preferences, respectively, under the framework of Markowitz mean-variance portfolio optimization. It would be interesting to consider a model-free RL approach where the robo-advisor has the freedom to learn and improve decisions beyond a pre-specified set of strategies or the Markowitz framework.

\paragraph{Sample Efficiency in Learning Trading Strategies.}
In recent years, sample complexity has been studied extensively  to understand modern reinforcement learning algorithms (see Sections \ref{sec:rl_basics}-\ref{sec:deep_value_based}). However, most RL algorithms still require a large number of samples to train a decent trading algorithm, which may exceed the amount of relevant available historical data. Financial time series are known to be non-stationary \cite{huang2003applications}, and hence historical data that are further away in time may not be helpful in training efficient learning algorithms for the current market environment. This leads to important questions of designing more sample-efficient RL algorithms for financial applications or developing good market simulators that could generate (unlimited) realistic market scenarios \cite{wiese2020quant}.

\paragraph{Transfer Learning and Cold Start for Learning New Assets.}
Financial institutions or individuals may change their baskets of assets to trade over time. Possible reasons may be that new assets (for example cooperative bonds) are issued from time to time or the investors may switch their interest from one sector to another. There are two interesting research directions related to this situation. When an investor has a good trading strategy, trained by an RL algorithm for one asset, how should they transfer the experience to train a trading algorithm for a ``similar'' asset with fewer samples? This is closely related to transfer learning \cite{torrey2010transfer,pan2009survey}. To the best of our knowledge, no study for financial applications has been carried out along this direction. Another question is the cold-start problem for newly issued assets. When we have very limited data for a new asset, how should we initialize an RL algorithm and learn a decent strategy using the limited available data and our experience (i.e., the trained RL algorithm or data) with other longstanding assets?

\paragraph{Acknowledgement} We thank Xuefeng Gao, Anran Hu, Xiao-Yang Liu, Wenpin Tang, Ziyi Xia, Zhuoran Yang,  Junzi Zhang and Zeyu Zheng  for helpful discussions and comments on this survey.

\bibliographystyle{siam}
\bibliography{references}
\end{document}